\documentclass{article}
\PassOptionsToPackage{numbers, compress, sort}{natbib}
\usepackage[final]{neurips_data_2024}

\usepackage[utf8]{inputenc} 
\usepackage[T1]{fontenc}    
\usepackage[hidelinks]{hyperref}       
\usepackage{booktabs}       
\usepackage{amsfonts}       
\usepackage{nicefrac}       
\usepackage{microtype}      
\usepackage[dvipsnames, table]{xcolor}
\usepackage{amssymb}
\usepackage{amsmath}
\usepackage{verbatim}
\usepackage{graphicx}
\usepackage{multirow}
\usepackage{booktabs}
\usepackage{natbib}
\usepackage{titlesec}
\usepackage[ruled,vlined]{algorithm2e}
\usepackage[utf8]{inputenc}
\usepackage{amssymb,amsmath,caption}
\usepackage[hang,flushmargin]{footmisc}
\usepackage{lipsum}
\usepackage{listings}
\usepackage{extarrows}
\usepackage{subfigure}
\usepackage{xspace}
\usepackage{graphicx}
\usepackage{colortbl} 
\usepackage{makecell}
\usepackage{ifthen}
\usepackage{xifthen}
\usepackage{wrapfig}
\usepackage{amsthm}
\usepackage{stmaryrd}
\usepackage{threeparttable}
\usepackage{enumitem}
\usepackage{multicol}
\usepackage{bbold}
\usepackage{tabularx}
\usepackage[framemethod=TikZ]{mdframed}

\usepackage{listings}
\newcommand{\hw}[1]{\textcolor{black}{#1}} 

\newcommand{\revision}[1]{\textcolor{black}{#1}}
\usepackage{courier}

\newcommand{\benchmarkt}{\textsc{STaRK}} 
\newcommand{\benchmarkh}{\textsc{STaRK }}
\newcommand{\amazont}{\textsc{STaRK-Amazon}}
\newcommand{\amazonh}{\textsc{STaRK-Amazon }}
\newcommand{\magt}{\textsc{STaRK-MAG}}
\newcommand{\magh}{\textsc{STaRK-MAG }}
\newcommand{\primekgt}{\textsc{STaRK-Prime}}
\newcommand{\primekgh}{\textsc{STaRK-Prime }}
\newcommand{\ie}{\textit{i.e., }}
\newcommand{\eg}{\textit{e.g., }}

\newcommand{\etc}{\textit{etc.}}

\newcommand{\xhdr}[1]{{\vspace{-2pt}\noindent\bfseries #1}.}

\newcommand{\tohide}[1]{}

\title{\benchmarkt: Benchmarking LLM Retrieval on Textual and Relational Knowledge Bases }
\author{%
  Shirley Wu$^{*\S}$,
   Shiyu Zhao$^{*\S}$, 
   Michihiro Yasunaga$^{\S}$, 
  Kexin Huang$^{\S}$, Kaidi Cao$^{\S}$, Qian Huang$^{\S}$, \\
  \textbf{Vassilis N. Ioannidis}$^{\dag}$, \textbf{Karthik Subbian}$^{\dag}$, \textbf{James Zou}$^{\ddag\S}$, \textbf{Jure Leskovec}$^{\ddag\S}$\\
  $^{\S}$Department of Computer Science, Stanford University\ \ \ $^{\dag}$Amazon\\
  {\textcolor{cyan}{\url{https://stark.stanford.edu/}}}
}

\begin{document}

\renewcommand{\thefootnote}{\fnsymbol{footnote}}
\footnotetext{$^{*\ddag}$Equal first-author / senior contribution. \\Correspondence: \texttt{\{shirwu, jamesz, jure\}@cs.stanford.edu}.}

\maketitle

\vspace{-10pt}
\begin{abstract}  
\label{sec:abs}
    

Answering real-world complex queries, such as complex product search, often requires accurate retrieval from semi-structured knowledge bases that involve blend of unstructured (\eg textual descriptions of products) and structured (\eg entity relations of products) information. 
However, many previous works studied textual and relational retrieval tasks as separate topics.
To address the gap, we develop \benchmarkt, a large-scale \underline{S}emi-structure retrieval benchmark on \underline{T}extual \underline{a}nd \underline{R}elational \underline{K}nowledge Bases.
Our benchmark covers three domains: product search, academic paper search, and queries in precision medicine.
We design a novel pipeline to synthesize realistic user queries that integrate diverse relational information and complex textual properties, together with their ground-truth answers (items). 
We conduct rigorous human evaluation to validate the quality of our synthesized queries. 
\hw{We further enhance the benchmark with high-quality human-generated queries to provide an authentic reference.}
\benchmarkh serves as a comprehensive testbed for evaluating the performance of retrieval systems driven by large language models (LLMs). Our experiments suggest that \benchmarkh presents significant challenges to the current retrieval and LLM systems, 
highlighting the need for more capable semi-structured retrieval systems. 

\end{abstract}

\begin{figure}[h]
    \centering
    \includegraphics[width=0.90\textwidth]{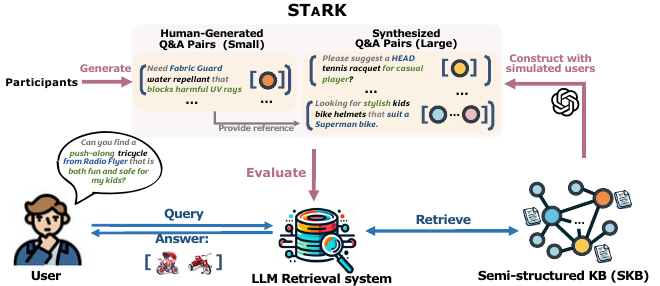}
    \caption{
    \benchmarkh features queries on Semi-structured Knowledge Base (SKB) with textual and relational knowledge, with node entities as ground-truth answers. \revision{\benchmarkh consists of synthesized queries simulating user interactions with a SKB and human-generated queries which provide an authentic reference. }It evaluates LLM retrieval systems' performance in providing accurate responses. 
    }
    \label{fig:task}
\end{figure}

\begin{table}
    \resizebox{1.00\textwidth}{!}{
    \begin{tabular}{lcl}
        \toprule
        &{Example query} & Title of ground-truth items(s) \\
        \midrule
        \hline
        \multirow{4}{*}{\amazont}& \cellcolor{cyan!5} {\textit{Looking for \textcolor{OliveGreen}{durable} \textcolor{MidnightBlue}{Dart World brand} {dart flights} that}} & \cellcolor{cyan!5}  {\small{<Amazon Standard Flights>}} \\
         & \cellcolor{cyan!5} {\textit{\textcolor{OliveGreen}{resist easy tearing}. Any recommendations?}}  & \cellcolor{cyan!5} {{\small{<Dart World Broken Glass Flight> (+12 more)}}}  \\
        \noalign{\vspace{0.75ex}}
        &\cellcolor{cyan!5} {\textit{What are recommended {scuba diving weights} \textcolor{OliveGreen}{for experienced divers} }} & \cellcolor{cyan!5} {{\small{<Sea Pearls Vinyl Coated Lace Thru Weight>}}}\\
        &\cellcolor{cyan!5} {\textit{that would \textcolor{MidnightBlue}{fit well with my Gorilla PRO XL waterproof bag}?}}& \cellcolor{cyan!5} {}\\
        \midrule
        \multirow{ 4}{*}{\magt}  & \cellcolor{yellow!10} {\textit{Search {publications} \textcolor{MidnightBlue}{by Hao-Sheng Zeng} on \textcolor{OliveGreen}{non-Markovian dynamics}.}} & \cellcolor{yellow!10} {{\small{<Distribution of non-Markovian intervals...>}}} \\
        &\cellcolor{yellow!10}{} &\cellcolor{yellow!10} {{\small{<Comparison between non-Markovian...>}}}\\
        \noalign{\vspace{0.75ex}}
        & \cellcolor{yellow!10}\textit{What are some \textcolor{OliveGreen}{nanofluid heat transfer} {research papers} published by} & \cellcolor{yellow!10} {{\small{<A Numerical Study on Convection Around A}}}\\
        &  \cellcolor{yellow!10}\textit{scholars \textcolor{MidnightBlue}{from Philadelphia University}?}  & \cellcolor{yellow!10} {{\small{Suqare Cylinder using AL2O3-H2O Nanofluid>}}}\\
        \midrule
        \multirow{6}{*}{\primekgt} &  \cellcolor{pink!20}\textit{Could you provide a list of \textcolor{OliveGreen}{investigational} {drugs} that } & \cellcolor{pink!20}{ \small{<(S)-3-phenyllactic Acid>}},\\
        &\cellcolor{pink!20}\textit{\textcolor{MidnightBlue}{interact with genes or proteins active in the epididymal region}?} & \cellcolor{pink!20}{ \small{<Anisomycin>}}, {\small{<Puromycin>}}\\
        \noalign{\vspace{0.75ex}}
        & \cellcolor{pink!20}\textit{Search for {diseases} \textcolor{MidnightBlue}{without known treatments} and induce} & \cellcolor{pink!20} {\small{<Intrahepatic Cholestasis>}}\\
        &\cellcolor{pink!20}\textit{\textcolor{OliveGreen}{pruritus in pregnant women}, potentially \textcolor{MidnightBlue}{associated with Autoimmune}.} & \cellcolor{pink!20}\\
        \noalign{\vspace{0.75ex}}
        & \cellcolor{pink!20}\textit{Please find {pathways} \textcolor{MidnightBlue}{involving the POLR3D gene} \textcolor{OliveGreen}{within nucleoplasm}.}& \cellcolor{pink!20}{\small{<RNA Polymerase III Chain Elongation>}} \\ 
        \noalign{\vspace{0.75ex}}
        & \cellcolor{pink!20}\textit{Which gene or protein \textcolor{MidnightBlue}{associated with lichen amyloidosis} can }& {\small{<OSMR>}},  {\small{<IL31RA>}}\cellcolor{pink!20}\\
        & \cellcolor{pink!20}\textit{\textcolor{OliveGreen}{bind interleukin-31 to activate the PI3K/AKT and MAPK pathways}?}& \cellcolor{pink!20}\\
        \bottomrule
    \end{tabular}
    }
    \vspace{2pt}
    \caption{\benchmarkh QA examples which involve semi-structured (\textcolor{MidnightBlue}{relational} and \textcolor{OliveGreen}{textual}) information. }
    \label{tab:benchmark_examples}
    \vspace{-15pt}
\end{table}

\section{Introduction}
\vspace{-6pt}
Natural-language queries are the primary form of how humans acquire information~\cite{hirschman2001natural, kaufmann2010evaluating,Jamil17}. 
For example, users on e-commerce sites wish to express complex information needs by combining free-form elements and constraints, such as ``\textit{Can you help me find a push-along tricycle from Radio Flyer that’s both fun and safe for my kid?}''\revision{ in product search. Medical scientists may ask questions like ``\textit{What disease is associated with the PNPLA8 gene and presents with hypotonia as a symptom?}''}.
Answering such queries is crucial for enhancing user experience, supporting informed decision-making, and preventing hallucination.

To answer such queries, the underlying knowledge can be represented in 
semi-structured knowledge bases (SKBs)~\cite{UniK-QA,mag,RyuJK14}, which integrate unstructured data, such as natural language descriptions and expressions (\eg description of the tricycle), with structured data, like entity interactions on knowledge graphs
(\eg a tricycle ``brand'' is Radio Flyer). 
This allows the SKBs to represent comprehensive knowledge in specific applications, making them indispensable in domains such as e-commerce~\cite{ups-amazon_review}, social media~\cite{MansmannRWS14}, and precision medicine~\cite{primekg, ogb, Johnson2021}.


\xhdr{Limitations of prior works} Prior works focused on either purely textual queries on unstructured knowledge~\cite{IzacardG21, realm, KarpukhinOMLWEC20, LeeCT19, YangXLLTXLL19, dunn2017searchqa, TriviaQA, yang2018hotpotqa} or structured SQL~\cite{zhongSeq2SQL2017, yu-etal-2018-spider, zhongSeq2SQL2017, yu-etal-2018-spider} or knowledge graph queries~\cite{BerantCFL13, YihCHG15, h2t, ComplexWebQuestions, yih-etal-2016-value,talmor-berant-2018-web,cao-etal-2022-kqa,GAO_FODOR_KIFER_2019,he2024gretriever,AsaiHHSX20}, which are limited in the span of knowledge and inadequate to study the complexities of retrieval on SKBs. Recently, large language models (LLMs) have demonstrated significant potential on information retrieval tasks~\cite{llm4ir, realm, rag, replug}. Nevertheless, it remains an open question of how effectively LLMs can be applied to the challenging retrieval tasks on SKBs.  Moreover, the existing works mainly focus mainly on general knowledge, \eg from Wikipedia. However, the knowledge may commonly come from private sources, requiring retrieval systems to \revision{operate} on private SKBs. Therefore, there is a gap of how current LLM retrieval systems handle the complex textual and relational requirements in queries that can involve private knowledge. 

\xhdr{Present work} 
To address this gap, we present a large-scale \underline{S}emi-structure retrieval benchmark on \underline{T}extual \underline{a}nd \underline{R}elational \underline{K}nowledge Bases (\benchmarkt) (Figure~\ref{fig:task}). The key technical challenge that we solve is how to accurately simulate user queries on SKBs. This difficulty arises from the interdependence of textual and relational information, which leads to challenges in precisely construct the ground-truth answers from millions of candidates. Additionally, ensuring that queries are useful and resembles real-world scenarios adds further complexity to the benchmarking process. 
 
We develop a novel pipeline that simulates user queries and constructs precise ground truth answers using three SKBs built from extensive texts and millions of entity relations from public sources.
We validate the quality of queries in our benchmark through detailed analysis and human evaluation, focusing on their naturalness, diversity, and practicality.
\revision{Furthermore, we incorporate 274 human-generated queries to compare with synthesized queries and enrich the testing scenarios.} With \benchmarkt, we delve deeper into retrieval tasks on SKBs, evaluate the capability of current retrieval systems, and provide insights for future advancement. 
Key features of \benchmarkh are:
\begin{itemize}[leftmargin=*]
    \vspace{-5pt}
    \item \textbf{Natural-sounding queries on SKBs (Table~\ref{tab:benchmark_examples}):} Queries in our benchmark incorporate rich relational information and complex textual properties. Additionally, these queries closely mirror the types of questions users would naturally ask in real-life scenarios, \eg with flexible query formats and possibly with additional contexts.
    \vspace{-1pt}
    \item \textbf{Context-specific reasoning:} The queries entail reasoning capabilities specific to the context. This includes the ability to infer customer interests, understand specialized field descriptions, and deduce relationships involving multiple subjects mentioned within the query.
    For example, the context  ``\textit{I had a dozen 2.5-inch Brybelly air hockey pucks, so I'm trying to find matching strikers.}'' entails the user's interest in looking for complementary products.
    Such reasoning capabilities are crucial for accurately interpreting and responding to the nuanced requirements of each query. 
    \vspace{-1pt}
    \item \textbf{Diverse domains:} 
    Our benchmark spans three knowledge bases\footnote{Explore the SKBs at \textcolor{cyan}{\url{https://stark.stanford.edu/skb_explorer.html}}} for applications including product recommendation, academic paper search, and precision medicine inquiries. \benchmarkh provides a comprehensive evaluation of retrieval systems across diverse contexts and domains. 
\end{itemize}

\vspace{-5pt}
We conduct extensive experiments on LLM retrieval systems, highlighting challenges in handling textual and relational data and latency on large-scale SKBs with millions of entities or relations. Finally, we offer insights into building more capable retrieval systems to handle real-world complexity.

\vspace{-5pt}
\section{Benchmarking 
Retrieval Tasks over Textual and Relational Knowledge}
\vspace{-5pt}

\begin{figure}[t]
    \centering
    \includegraphics[width=0.98\textwidth]{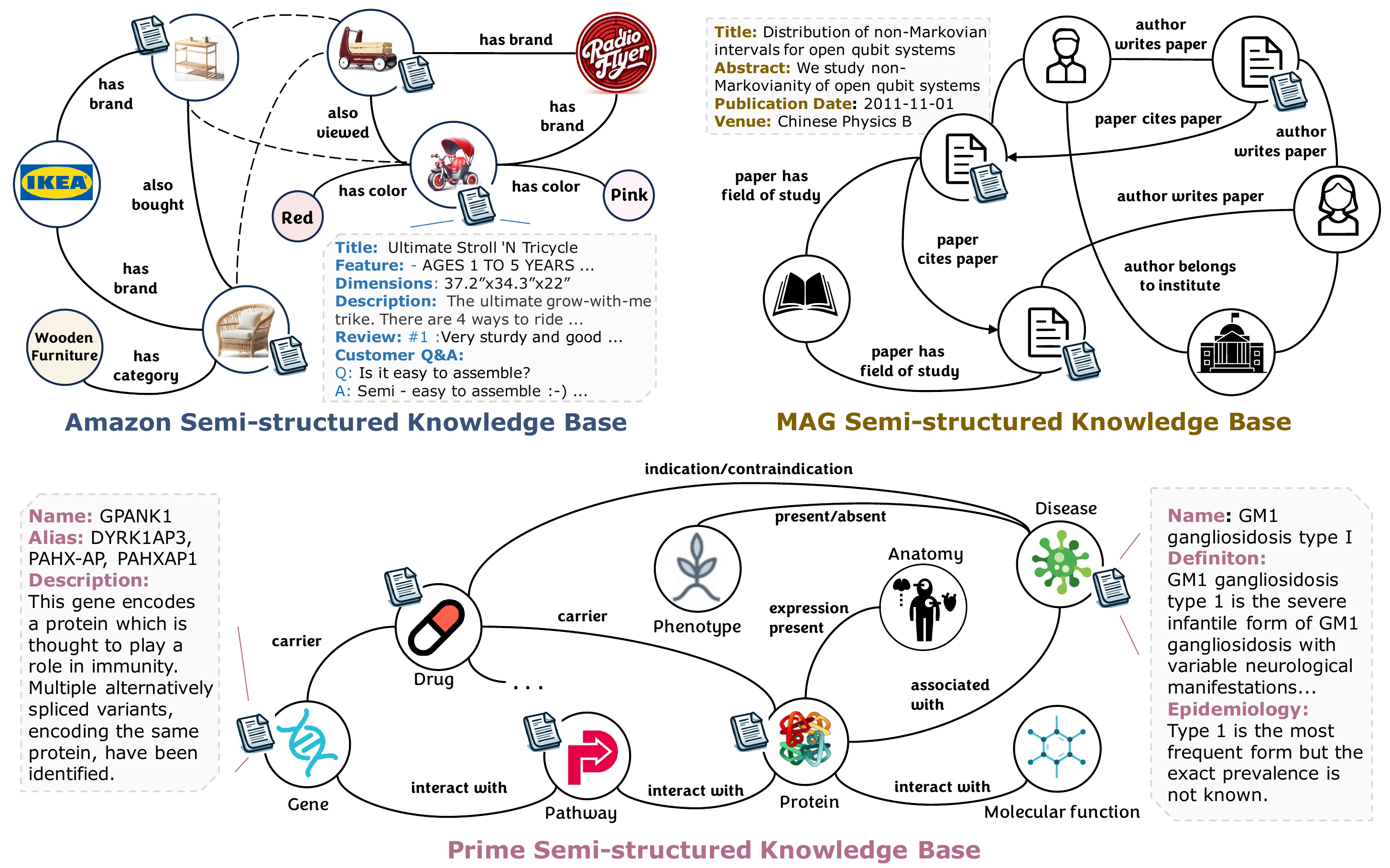}
    \vspace{-5pt}
    \caption{Demonstration of Semi-structured Knowledge Bases, 
    where each knowledge base combines both textual and relational information in a complex way, making the retrieval tasks challenging.
    }
    \vspace{-15pt}
    \label{fig:semistructure}
\end{figure}

\subsection{\revision{Problem Definition}}
\label{sec:def}
\vspace{-3pt}
We are given a Semi-Structured Knowledge Base (SKB), which consists of a knowledge graph \( G \) and a collection of free text documents  $D$. 
Formally, let \( G = (V, E) \) be the knowledge graph, where \( V \) is the set of nodes and \( E \subseteq V \times V \) is the set of edges representing relationships between nodes. \( D = \bigcup_{i \in V} D_i \) be the collection of free-form text documents associated with the nodes, where \( D_i \) is the set of documents associated with node \( i \). For example, the product knowledge graph in e-commerce can capture relationships between products and brands/colors/categories, and the corresponding text documents include product descriptions, reviews, \etc

We define the tasks on our benchmark datasets as follows: 
Given the knowledge graph \( G = (V, E) \), a collection of free text documents \( D \), and a query \( Q \), the output is a set of nodes \( A \subseteq V \) such that for each node \( i \in A \), it satisfies the relational requirements imposed by the structure of \( G \) as specified in \( Q \), and the associated documents \( D_i \) satisfy the textual requirements specified in \( Q \). 


\vspace{-3pt}
\subsection{Semi-structured Knowledge Bases (SKBs)}
\label{sec:skb}
\vspace{-3pt}

\begin{table}[t]
    \centering
    \caption{Data statistics of our constructed semi-structured knowledge bases}
    \resizebox{0.92\textwidth}{!}{
    \begin{tabular}{l|ccrrrr}
        \toprule
        \multirow{ 2}{*}{} & \textbf{\#entity} & \textbf{\#relation} & \textbf{avg. } & \multirow{ 2}{*}{\textbf{\#entities}} & \multirow{ 2}{*}{\textbf{\#relations}} & \multirow{ 2}{*}{\textbf{\#tokens}} \\
        & \textbf{types} &\textbf{types} & \textbf{degree}\\
        \midrule
        \amazont & 4 & 5  & 18.2 & 1,035,542 & 9,443,802 & 592,067,882\\
        \magt & 4 & 4  & 43.5& 1,872,968 & 39,802,116 & 212,602,571\\
        \primekgt & 10& 18 & 125.2 & 129,375 & 8,100,498  & 31,844,769\\
        \bottomrule
    \end{tabular}
    }
    \vspace{-15pt}
    \label{tab:statistics}
\end{table}

As shown Figure~\ref{fig:semistructure}, we construct three large-scale SKBs with the relational and textual information with each entity. 
See Table~\ref{tab:statistics} for the basic data statistics and Appendix~\ref{app:skb} for details.

\xhdr{Amazon Semi-structured Knowledge Base}  
The SKB features four entity types: \texttt{product}, \texttt{brand}, \texttt{color}, and \texttt{category}, and five relation types:  \texttt{also\_bought, also\_viewed} between \texttt{product} entities, and
\texttt{has\_brand/color/category} associated with the products. 
We derive the textual information of \texttt{product} nodes by combining Amazon Product Reviews~\cite{ups-amazon_review} with Amazon Q\&A Data~\cite{image_amazon_qa}. 
This provides a rich amount of texts, including product descriptions and customer reviews. 
For other entities, we extract their names or titles as the textual attributes. Amazon SKB features an extensive textual data largely contributed from customer reviews and Q\&A.

\xhdr{MAG Semi-structured Knowledge Base} 
This SKB includes node entities of \texttt{paper}, \texttt{author}, \texttt{institute}, and \texttt{field\_of\_study}. We derive its relational structure by extracting a subgraph from obgn-mag~\cite{hu2021open}, which contains shared paper nodes with obgn-papers100M~\cite{hu2021open} and all non-paper nodes. We filter out non-English language papers as we only consider single-lingual queries. 
The paper documents include their titles and abstracts. Additionally, we integrating details from the Microsoft Academic Graph database (version 2019-03-22)~\cite{mag, mag_data}, providing extra textual information like paper venue, author and institution names. 
This SKB demonstrates a large number of relations associate with \texttt{paper} nodes, especially on citation and authorship relations.

\xhdr{Prime Semi-structured Knowledge Base} We leverage the exisiting knowledge graph PrimeKG~\cite{primekg} which contains ten entity types including \texttt{disease, gene/protein}, and eighteen relation types, such as \texttt{associated\_with, indication}.
Compared to the Amazon and MAG SKBs, Prime SKB is denser and features a greater variety of relation types. While PrimeKG provides text information on \texttt{disease} and \texttt{drug} entities, 
we integrate the data from multiple databases for \texttt{gene/protein} and \texttt{pathway} entities such as genomic position, gene activity summary and pathway orthologous event.

\subsection{Retrieval Tasks on Semi-structured Knowledge Bases}
\label{sec:star}
Our retrieval benchmark (Table~\ref{tab:benchmark_stats}) consists of three novel retrieval-based question-answering datasets, each comprising synthesized train/val/test sets with $9$k to $14$k queries in total and a high-quality human-generate query set.
The queries synthesize relational and textual knowledge, mirroring real-world queries in terms of natural-sounding property and flexible formats.


\xhdr{\amazont} 
The task aims at product recommendation, with a notable 68\% of the synthesized queries yielding more than one ground truth answer. The dataset prioritizes customer-oriented criteria, highlighting textual elements such as product quality, functionality, and style. Moreover, it incorporate single-hop relational aspects (Appendix~\ref{app:amazon}) into the queries, including brand, category, and product connections (\eg complementary or substitute items). The queries are framed in conversation-like formats, enriching the context and enhancing the dataset's relevance to real-world scenarios.

\xhdr{\magt} 
Beyond the single-hop relational requirements in \amazont, \magh emphasizes the fusion between the textual requirements with multi-hop queries for precise academic paper search. For example, ``Are there any papers from King's College London'' highlights the metapath (\texttt{institution $\rightarrow$ author $\rightarrow$ paper}) on the relational structure. 
We designed three single-hop and four multi-hop relational query templates (Appendix~\ref{app:mag}). The textual aspects focus on the paper's topic, methodology, and contribution \etc

\xhdr{\primekgt} 
The task is to answer complex biomedicine inquiries. 
For synthesized queries, we developed 28 multi-hop query templates (Appendix~\ref{app:prime}) to cover various relation types and ensure their practical relevance. 
For example, the template ``What is the drug that targets genes or proteins in <anatomy>?'' aids precision medicine by identifying treatments targeted to specific anatomical areas.
For \texttt{drug, disease, gene/protein}, and \texttt{pathway} entities, the queries are a hybrid of relational and textual requirements. 
For entities such as \texttt{effect/phenotype}, the queries rely solely on relational data due to limited textual information. 
We exhibit three distinct user roles – medical scientist, doctor, and patient – for generating queries about drug and disease, which diversify the language to comprehensively evaluate the retrieval systems.

\begin{table}[t]
    \centering
    \caption{Statistics on the \benchmarkh benchmark datasets.}
    \resizebox{0.9\textwidth}{!}{
    \begin{tabular}{ll|rrrr}
        \toprule
        & & \multirow{2}{*}{\textbf{ \#queries}}  & \textbf{ \#queries w/} & \textbf{average} & \multirow{2}{*}{\textbf{train / val / test}}\\
        & & & \textbf{multiple answers}&  \textbf{\#answers} & \\
        \midrule
        \multirow{3}{*}{
        \makecell{Synthesized\\
        (Sec~\ref{sec:construction}, \ref{sec:analysis})}} & {\small \amazont} & 9,100 & 7,082  & 17.99& {0.65 / 0.17 / 0.18}\\
        & {\small \magt} &13,323 & 6,872 & 2.78 & {0.60 / 0.20 / 0.20}\\
        & {\small \primekgt} & 11,204 & 4,188 & 2.56 & {0.55 / 0.20 / 0.25} \\
        \midrule
        \multirow{3}{*}{\makecell{Human-generated\\
        (Sec~\ref{sec:human_generated})}} 
        & {\small \amazont} & 81 & 64 & 19.50 & \multirow{3}{*}{For testing only}\\
        & {\small \magt} & 84 &34 & 3.26 & \\
        & {\small \primekgt} & 98 & 41 & 2.77 & \\
        \bottomrule
        \end{tabular}
        }
    \vspace{-5pt}
    \label{tab:benchmark_stats}
\end{table}
\begin{figure}[t]
    \centering
    \includegraphics[width=0.95\textwidth]{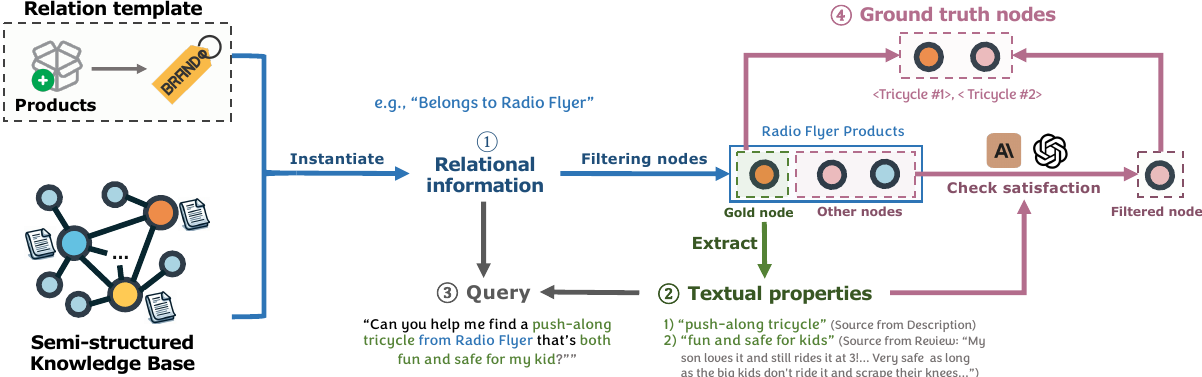}
    \vspace{-3pt}
    \caption{
    The construct pipeline to generate our semi-structured retrieval datasets.
    }
    \vspace{-10pt}
    \label{fig:construction}
\end{figure}

\vspace{-2pt}
\subsection{Benchmark Construction: Synthesized Queries} 
\label{sec:construction}
\vspace{-2pt}
In Figure~\ref{fig:construction}, we present a novel pipeline that synthesizes the SKB queries and automatically generates the ground truth answers. 
The key idea is to entangle relational and textual information during synthesis and disentangle them during answer filtering. 
It involves four steps as follows:


\begin{itemize}[leftmargin=*]
\vspace{-2pt}
\item \textbf{1) Sample Relational Requirements:} For each query, we sample a practical relation template constructed with expert/domain knowledge, \eg ``\texttt{(a product) belongs to <brand>}'' and ground it with sampled entities (\ie a specific brand), \eg ``belongs to Radio Flyer''. 
This relational requirement yields a set of candidate entities, \ie products belonging to Radio Flyer. 
\vspace{-2pt}
\item \textbf{2) Extracting Textual Properties:} \revision{
We randomly sample a candidate entity from the first step, referred to as the \textit{gold answer}, from which LLMs extract properties that align with the interests of specific roles (\eg customers, researchers, or doctors) in its textual document}.  In Figure~\ref{fig:construction}, we extract multiple properties about the functionality and user experience from a Radio Flyer product. 
\vspace{-2pt}
\item \textbf{3) Combining Textual and Relational Information:} 
We use two LLMs to synthesize queries from textual properties and relational requirements, enhancing diversity and reducing bias arise from relying on a single LLM. 
The first LLM focuses on generating natural, role-specific, and style-consistent (\eg ArXiv searches) queries. The second LLM enriches the context and rephrases queries, which poses the need for advanced reasoning to comprehend them under complex contexts.

\vspace{-2pt}
\item \textbf{4) Filtering Additional Answers:} 
Finally, we employ multiple LLMs to verify if the candidates from the first step meet the extracted textual properties. Only candidates passing all LLM verifications are included in the final ground truth set. To assess the precision of this filtering mechanism, we compute the average ratios for the gold answers to be verified, which are 86.6\%, 98.9\%, and 92.3\% on the three datasets, highlighting our efficacy in yielding high-quality ground truth answers.

\vspace{-2pt}
\end{itemize}

This dataset construction pipeline is automatic, efficient, and broadly applicable to the SKBs in our formulation.  
\revision{We include all of the prompts and the LLMs versions in the above steps in Appendix~\ref{app:prompts}.}

\begin{figure}[t]
    \centering
    \includegraphics[width=1\textwidth]{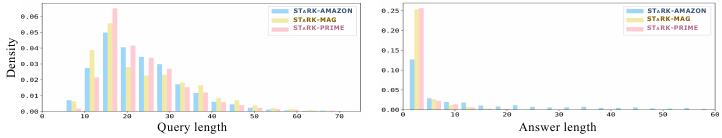}
    \vspace{-15pt}
    \caption{Distribution of query and answer lengths on \benchmarkh datasets.}
    \vspace{-5pt}
    \label{fig:length_dist}
\end{figure}

\begin{figure}[t]
    \begin{minipage}[t]{0.5\textwidth}
        \centering
      \captionof{table}{Query diversity measurement on \benchmarkt. \revision{See Appendix~\ref{app:math} for the metric definition.}}
      \vspace{5pt}
        \resizebox{1\textwidth}{!}{
        \begin{tabular}{l|cc}
            \toprule
            & {\small Shannon Entropy} & {\small Type-Token Ratio} \\
            \midrule
            \small \amazonh & 10.39 & 0.179 \\
            \small \magh & 10.25 & 0.180 \\
            \small \primekgh & 9.63 & 0.143 \\
            \midrule
            \small Reference article & 10.44 & 0.261\\
            \bottomrule
        \end{tabular}
        }
        \label{tab:q_diversity}
    \end{minipage}%
    \hfill 
    \begin{minipage}[t]{0.47\textwidth}
        \caption{Average relative composition of relational vs. textual information.}
        \vspace{-5pt}
        \includegraphics[width=\textwidth]{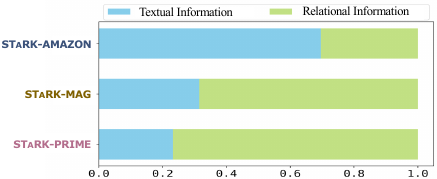}
        \label{fig:ratio}
    \end{minipage}
    \vspace{-20pt}
\end{figure}

\subsection{Synthesized Data Distribution Analysis and Human Evaluation}
\label{sec:analysis}

\begin{itemize}[leftmargin=*]
    \item \xhdr{Query and Answer Length} 
    Query length (in words) reflects the amount of user-provided context information, while the number of answers indicates query ambiguity/concreteness.
    Figure~\ref{fig:length_dist} shows similar query length distributions across the datasets, with most queries around 16 words. Longer queries (up to 50 words) often mention other entities or provide detailed context. Notably, 
    \amazonh has a significant long-tail pattern, with about 22\% of the answers have more than 30 entities, reflecting diverse e-commerce recommendations and ambiguous user queries.
     
    \item \xhdr{Query Diversity} 
    A diverse set of queries poses challenges for broader applicability to meet varying user demands. We measure query diversity using Shannon Entropy for word distribution and Type-Token Ratio (TTR) for unique words. Higher values indicate greater lexical diversity. Table~\ref{tab:q_diversity} shows high Shannon Entropy and steady TTR across all datasets. For reference, we compute these metrics for the Wikipedia page of Barack Obama\footnote{\url{https://en.wikipedia.org/wiki/Barack\_Obama}}. 
    
    \item \xhdr{Proportionality of Relational \textit{vs.}Textual Information}
    Our benchmark queries feature the composition of textual and relational information. To understand the distribution of information types, we calculate the average ratio of relational to textual requirements by word count in the queries across each dataset.  
    Note that the ratios do not directly reflect their importance in determining final answers. Figure~\ref{fig:ratio} shows varying ratios, which highlights different emphases on textual versus relational requirements and challenges retrieval systems to adapt to different distributions.

\end{itemize}

\xhdr{Human evaluation} 
We qualitatively assess sampled queries from our benchmark for naturalness (resembling natural conversation), diversity (covering various question structures and complexities), and practicality (relevance to real-world situations) with 63 participants. Evaluation results, converted from a 5-point Likert-like scale to a positive/tie/negative scale, show positive and non-negative rates in Table~\ref{tab:human} (Appendix~\ref{app:exp}). On average, 94.1\%, 85.3\%, and 89.4\% of participants rated the queries neutral or above in naturalness, diversity, and practicality, respectively. These results validate the quality of our benchmark and its potential for diverse and realistic retrieval tasks.

\subsection{\revision{Benchmark Construction: Human-Generated Queries}}
\label{sec:human_generated}
To enhance our benchmark's practical relevance, we engaged 31 participants (22 native English speakers) to generate 263 queries across three SKBs following the detailed instructions (Appendix~\ref{app:instruction}) along with our interactive platform. We manually verified and filtered the ground truth answers to ensure the answer correctness. Table~\ref{tab:benchmark_stats} shows the statistics of the human-generated datasets. Finally, we analyzed the commonalities and differences between synthesized and human-generated queries.


\begin{table}[h]
\centering
\vspace{-13pt}
\caption{Comparison of Human-generated and Synthesized Queries}
\label{tab:comparison_queries}
\begin{tabularx}{\textwidth}{|c|X|X|}
\hline
\footnotesize \textbf{Query Type} & \footnotesize \textbf{Human-generated Query} & \footnotesize \textbf{Synthesized Query} \\
\hline
\footnotesize Short and Direct & \footnotesize \textit{Red sweatshirt for proud Montreal Canadiens} & \footnotesize \textit{Suggestions for a Suunto bike mount?} \\
\hline
\makecell{\footnotesize Specific \\ \footnotesize Author \& Field}  & \textit{\footnotesize \makecell[l]{Find me papers that discuss improving\\ condenser performance authored by \\Stojan Hrnjak}} & \textit{\footnotesize \makecell[l]{Show me papers by Seung-Hyeok Kye that \\discuss separability criteria.}} \\
\hline
\footnotesize \multirow{6}{*}{Complex Context} & \textit{\footnotesize Help me. I am trying to diagnose a patient with persistent joint pain, and I suspect a condition where the bone is dying due to compromised blood supply, often linked to factors like steroid use, ... what's the name of this \textbf{sneaky bone-killing culprit}?} & \textit{\footnotesize I'm experiencing joint pain accompanied by swelling... I'm concerned about medications aggravating my fuzzy eyesight and potential blood clotting complications. Could you recommend treatments while minimizing these side effects?} \\
\hline
\end{tabularx}
\vspace{-6pt}
\end{table}

\xhdr{Commonality}
The number of answers of synthesized and human-generated queries are comparable, indicating a similar level of query ambiguity. Moreover, we observe that most styles of human-generated queries are covered in the synthesized dataset. For example, Table~\ref{tab:comparison_queries} highlights their similarities in short product queries, specific author/field inquiries, and complex contextual queries.

\xhdr{Difference}
We find that human-generated queries often exhibit more unique expressions compared to synthesized ones, such as "\textit{Give me a \textbf{fat cross} and road tire that works with my Diamondback bicycle tube}" and "\textbf{\textit{this sneaky bone-killing culprit}}". This discovery suggests a future direction for our benchmark to incorporate modern and dynamic language nuances.

\vspace{-2pt}
\section{Experiments}
\label{sec:exp}
\vspace{-2pt}

\subsection{Baseline Retrieval Models and Evaluation Metrics}

We extensively evaluate five classes of retrieval models described below.
\vspace{-5pt}
\begin{itemize}[leftmargin=*]
    \vspace{-1pt}   
    \item \textbf{Sparse Retriever}: \textbf{BM25}~\cite{bm25} is a traditional yet powerful sparse retrieval method based on term frequency-inverse document frequency (TF-IDF). It computes relevance scores by considering the frequency of query terms in documents, adjusted for term rarity and document length.
    
    \vspace{-1pt}   
    \item \textbf{Small Dense Retrievers}: \textbf{DPR}~\cite{dpr}, \textbf{ANCE}~\cite{ance}, and \textbf{QAGNN}~\cite{qagnn}. These compact models generate dense embeddings for both queries and documents, computing retrieval scores based on embedding similarities. They serve as baselines for comparison with LLM-based dense retrievers.
    
    \vspace{-1pt}   
    \item \textbf{LLM-based Dense Retrievers}: \textbf{text-embedding-ada-002 (abbrev. ada-002)}~\cite{oai}, \textbf{voyage-large-2-instruct (abbrev. voyage-l2-instruct)}~\cite{voyage}, \textbf{LLM2Vec-Meta-Llama-3-8B-Instruct-mntp (abbrev. LLM2Vec)}~\cite{llm2vec}, and \textbf{GritLM-7b}~\cite{GritLM}. These models leverage LLMs to generate dense embeddings that are more contextually expressive.
    
    \vspace{-1pt}   
    \item \textbf{Multivector Retrievers}: \textbf{multi-ada-002}~\cite{oai} and \textbf{ColBERTv2}~\cite{ColBERTv2}. Beyond ada-002 which represents a document as an embedding, \textbf{multi-ada-002} splits each document into overlapping chunks and embeds them using the same encoder as the query. Similarity scores between the query and chunks are aggregated using the average of the top-3 similarities, which we found to perform best. \textbf{ColBERTv2} represents each document as multiple token-level embeddings for fine-grained matching, capturing richer semantic information.
    
    \vspace{-1pt}   
    \item \textbf{LLM Rerankers}: \textbf{Claude3} and \textbf{GPT-4} rerankers~\cite{abs-2306-04757, abs-2310-14122}. These models improve the precision of top-$k$ ada-002 results by reranking them using large language models. We employ GPT-4-turbo (\texttt{gpt-4-1106-preview}) and Claude3 (\texttt{claude-3-opus}), setting $k=20$ for synthesized queries and $k=10$ for human-generated queries. Given a query, the LLMs assign a satisfaction score from 0 to 1 to each candidate entity based on textual and relational information. Due to high computational costs, we evaluate these rerankers on a random 10\% sample of test queries.
\end{itemize}

The performance of these models are measured using standard retrieval metrics below.




\vspace{-5pt}
\begin{itemize}[leftmargin=*]

    \vspace{-1pt}   
    \item \textbf{Hit@$k$} assesses whether the correct item is among the top-$k$ results from the model. We used $k=1$ and $k=5$ for evaluation. At $k=1$, it evaluates the accuracy of the top recommendation; at $k=5$, it examines the model's precision in a wider recommendation set.  

    \vspace{-1pt}   
    \item \textbf{Recall@$k$} measures the proportion of relevant items in the top-$k$ results. For synthesized queries, $k=20$ is used, as the answer length of all of the queries in our benchmarks are equal or smaller then $20$. 
    This metric offers insight into the model's ability to identify all relevant items, particularly in scenarios where missing any could be critical. 

    \vspace{-1pt}   
    \item \textbf{Mean Reciprocal Rank (MRR)} is a statistic for evaluating the average effectiveness of a predictive model. It calculates the reciprocal of the rank at which the first relevant item appears in the list of predictions. 
    This metric emphasizes the importance of the rank of the first correct answer, which is crucial in many practical applications where the first correct answer is often the most impactful.
\end{itemize}

\subsection{Results and Analysis}
\vspace{-2pt}

\begin{table}[t]
    \centering
    \vspace{-10pt}
    \caption{Testing results on \benchmarkt-Syn(thesized).}
    \resizebox{1.0\textwidth}{!}{
    \begin{tabular}{r|rrrr|rrrr|rrrr}
        \toprule
        & \multicolumn{4}{c|}{\amazont} & \multicolumn{4}{c|}{\magt} & \multicolumn{4}{c}{\primekgt} \\
         & Hit@1 & Hit@5 & R@20 & MRR & Hit@1 & Hit@5 & R@20 & MRR & Hit@1 & Hit@5 & R@20 & MRR \\
        \hline
        \rowcolor{cyan!10} \multicolumn{13}{c}{\textbf{Full Testing Dataset}} \\
        \hline
        BM25 & 44.94 & \textbf{67.42} & 53.77 & 55.30 & 25.85 & 45.25 & 45.69 & 34.91 & 12.75 & 27.92 & 31.25 & 19.84 \\
        DPR (roberta) & 15.29 & 47.93 & 44.49 & 30.20 & 10.51 & 35.23 & 42.11 & 21.34 & 4.46 & 21.85 & 30.13 & 12.38 \\
        ANCE (roberta) & 30.96 & 51.06 & 41.95 & 40.66 & 21.96 & 36.50 & 35.32 & 29.14 & 6.53 & 15.67 & 16.52 & 11.05 \\
        
        QAGNN (roberta) & 26.56 & 50.01 & 52.05 & 37.75 & 12.88 & 39.01 & 46.97 & 29.12 & 8.85 & 21.35 & 29.63 & 14.73 \\
        ada-002 & 39.16 & 62.73 & 53.29 & 50.35 & 29.08 & 49.61 & 48.36 & 38.62 & 12.63 & 31.49 & 36.00 & 21.41 \\
        voyage-l2-instruct & 40.93 & 64.37 & 54.28 & 51.60 & 30.06 &  50.58 & \textbf{50.49} & 39.66 & 10.85 & 30.23 & 37.83 & 19.99 \\ 
        LLM2Vec & 21.74 & 41.65 & 33.22 & 31.47 & 18.01 & 34.85 & 35.46 & 26.10 & 10.10 & 22.49 & 26.34 & 16.12 \\
        GritLM-7b & 42.08 & 66.87 & \textbf{56.52} & 53.46 & \textbf{37.90} & \textbf{56.74} & 46.40 & \textbf{47.25} & \textbf{15.57} & 33.42 & \textbf{39.09} & \textbf{24.11} \\
        multi-ada-002 & 40.07 & 64.98 & 55.12 & 51.55 & 25.92 & 50.43 & 50.80 & 36.94 & 15.10 & \textbf{33.56} & 38.05 & 23.49 \\
        ColBERTv2 & \textbf{46.10} & 66.02 & 53.44 & \textbf{55.51} & 31.18 & 46.42 & 43.94 & 38.39 & 11.75 & 23.85 & 25.04 & 17.39 \\
        \hline
        \rowcolor{green!8} \multicolumn{13}{c}{\textbf{Random 10\% Sample}} \\
        \hline
        BM25 & 42.68 & 67.07 & 54.48 & 54.02 & 27.81 & 45.48 & 44.59 & 35.97 & 13.93 & 31.07 & 32.84 & 21.68 \\
        DPR (roberta) & 16.46 & 50.00 & 42.15 & 30.20 & 11.65 & 36.84 & 42.30 & 21.82 & 5.00 & 23.57 & 30.50 & 13.50 \\
        ANCE (roberta) & 30.09 & 49.27 & 41.91 & 39.30 & 22.89 & 37.26 &44.16 & 30.00 & 6.78 & 16.15 & 17.07 & 11.42 \\
        QAGNN (roberta) & 25.00 & 48.17 & 51.65 & 36.87 & 12.03 & 37.97 & 47.98 & 28.70 & 7.14 & 17.14 & 32.95 & 16.27 \\
        ada-002 & 39.02 & 64.02 & 49.30 & 50.32 & 28.20 & 52.63 & 49.25 & 38.55 & 15.36 & 31.07 & 37.88 & 23.50 \\
        voyage-l2-instruct & 43.29 & 67.68 & 56.04 & 54.20 & 34.59 & 50.75 & 50.75 & 42.90 & 12.14 & 31.42 & 37.34 & 21.23 \\
        LLM2Vec & 18.90 & 37.80 & 34.73 & 28.76 & 19.17 & 33.46 & 29.85 & 26.06 & 9.29 & 20.7 & 25.54 & 15.00 \\
        GritLM-7b & 43.29 & \textbf{71.34} & \textbf{56.14} & 55.07 & 38.35 & \textbf{58.64} & 46.38 & 48.25 & 16.79 & 34.29 & 41.11 & 24.99 \\
        multi-ada-002 & 40.85 & 62.80 & 52.47 & 51.54 & 25.56 & 50.37 & \textbf{53.03} & 36.82 & 15.36 & 32.86 & \textbf{40.99} & 23.70 \\
        ColBERTv2 & 44.31 & 65.24 & 51.00 & 55.07 & 31.58 & 47.36 & 45.72 & 38.98 & 15.00 & 26.07 & 27.78 & 19.98 \\
        Claude3 Reranker & \textbf{45.49} & 71.13 & 53.77 & \textbf{55.91} & 36.54 & 53.17 & 48.36 & 44.15 & 17.79 & 36.90 & 35.57 & 26.27 \\
        GPT4 Reranker & 44.79 & 71.17 & 55.35 & 55.69 & \textbf{40.90} & 58.18 & 48.60 & \textbf{49.00} & \textbf{18.28} & \textbf{37.28} & 34.05 & \textbf{26.55} \\
        \bottomrule
    \end{tabular}
    }
    \label{tab:results}
\end{table}

\begin{table}[t]
    \centering
    \vspace{-10pt}
    \caption{Testing results on \benchmarkt-Human(-Generated).}
    \resizebox{1.0\textwidth}{!}{
    \begin{tabular}{r|rrrr|rrrr|rrrr}
        \toprule
        & \multicolumn{4}{c|}{\textbf{\amazont}} & \multicolumn{4}{c|}{\textbf{\magt}} & \multicolumn{4}{c}{\textbf{\primekgt}} \\
        Method & Hit@1 & Hit@5 & R@20 & MRR & Hit@1 & Hit@5 & R@20 & MRR & Hit@1 & Hit@5 & R@20 & MRR \\
        \midrule
        BM25 & 27.16 & 51.85 & 29.23 & 18.79 & 32.14 & 41.67 & 32.46 & 37.42 & 22.45 & 41.84 & 42.32 & 30.37 \\
        DPR (roberta) & 16.05 & 39.51 & 15.23 & 27.21 & 4.72 & 9.52 & 25.00 & 7.90 & 2.04 & 9.18 & 10.69 & 7.05 \\
        ANCE (roberta) &25.93 & 54.32 & 23.69 & 37.12 & 25.00 & 30.95 & 27.24 & 27.98 & 7.14 & 13.27 & 11.72 & 10.07 \\
        QAGNN (roberta) & 22.22 & 49.38 & 21.54 & 31.33 & 20.24 & 26.19 & 28.76 & 25.53 & 6.12 & 13.27 & 17.62 & 9.39 \\
        ada-002 & 39.50 & 64.19 & 35.46 & 52.65 & 28.57 & 41.67 & 35.95 & 35.81 & 17.35 & 34.69 & 41.09 & 26.35 \\
        voyage-l2-instruct & 35.80 & 62.96 & 33.01 & 47.84 & 22.62 & 36.90 & 32.44 & 29.68 & 16.33 & 32.65 & 39.01 & 24.33 \\
        LLM2Vec & 29.63 & 46.91 & 21.21 & 38.61 & 16.67 & 28.57 & 21.74 & 21.59 & 9.18 & 21.43 & 26.77 & 15.24 \\
        GritLM-7b & 40.74 & 71.60 & 36.30 & 53.21 & 34.52 & 44.04 & 34.57 & 38.72 & 25.51 & 41.84 & \textbf{48.10} & 34.28 \\ 
        multi-ada-002 & 46.91 & 72.84 & \textbf{40.22} & 58.74 & 23.81 & 41.67 & \textbf{39.85} & 31.43 & 24.49 & 39.80 & 47.21 & 32.98 \\
        ColBERTv2 & 33.33 & 55.56 & 29.03 & 43.77 & 33.33 & 36.90 & 30.50 & 35.97 & 15.31 & 26.53 & 25.56 & 19.67 \\
        Claude3 Reranker & \textbf{53.09} & 74.07 & 35.46 & \textbf{62.11} & \textbf{38.10} & 45.24 & 35.95 & \textbf{42.00} & \textbf{28.57} & \textbf{46.94} & 41.61 & \textbf{36.32} \\
        GPT4 Reranker & 50.62 & \textbf{75.31} & 35.46 & 61.06 & 36.90 & \textbf{46.43} & 35.95 & 40.65 & \textbf{28.57} & 44.90 & 41.61 & 34.82 \\
        \bottomrule
    \end{tabular}
    }
    \label{tab:results-human}
    \vspace{-10pt}
\end{table}

\textbf{Results on synthesized queries.} 
Table~\ref{tab:results} presents the results on both the full synthesized test sets and random 10\% samples from these sets. In both cases, \textbf{BM25}, despite its simplicity, proves to be a strong baseline, outperforming the dense retrieval models such as \textbf{ANCE}. We observe that finetuned \textbf{DPR} and \textbf{QAGNN}, exhibit insufficient performance. This underperformance is likely due to their relatively small model sizes and the risk of overfitting during training. These issues present challenges in effectively training the models on SKBs, where the entity documents can be hard to differentiate without capturing detailed information.

Among the larger models, \textbf{ada-002} benefits from superior pretrained embeddings and significantly outperforms \textbf{LLM2Vec} by a large margin. \textbf{GritLM-7b} delivers excellent performance, surpassing the ada-002 model overall. In contrast, LLM2Vec underperforms due to its limited context length, which is insufficient for encoding the lengthy documents in the SKBs. For multivector retrievers, we found that \textbf{multi-ada-002} generally outperforms ada-002, indicating that using multiple vectors per document enhances retrieval effectiveness. Similarly, fine-grained representation allows \textbf{ColBERTv2} to capture subtle semantic nuances between queries and documents, leading to largely improved retrieval accuracy. 

However, both GritLM-7b and ColBERTv2 generally underperform compared to the rerankers on the random split, especially in terms of Hit@k metrics. This suggests that while these dense retriever models effectively capture semantic information, they may not fully grasp the nuanced relevance judgments required for top-tier retrieval performance. The rerankers, utilizing powerful LLMs like \textbf{GPT-4 (gpt-4-1106-preview)} and \textbf{Claude3 (claude-3-opus)}, excel by re-evaluating the top candidates and assigning satisfaction scores based on a deeper understanding of the query and document content. This process allows them to better discern subtle contextual cues and relational information that dense retrievers might overlook. Consequently, LLM rerankers enhance retrieval precision at the top ranks.

Finally, regardless of the higher computational costs of the rerankers, their performance remains suboptimal. For instance, the Hit@1 scores for the GPT-4 reranker are only about 18\% on \primekgt and 41\% on \magt, indicating that the top-ranked answers are frequently incorrect. Similarly, the Recall@20 metrics are below 60\% across all datasets, with the GPT-4 reranker achieving Recall@20 scores of 55\% on \amazont, 49\% on \magt, and 34\% on \primekgt. This suggests that the ranking results miss a significant portion of relevant answers. The MRR scores are also relatively low, especially for \primekgt, where the GPT-4 reranker attains an MRR of only around 27\%. 

The insufficient performances may be attributed to the complexity and diversity of queries in SKBs, where nuanced understanding and detailed contextual information are crucial.
These findings highlight significant room for improvement in the ranking process.

\xhdr{Results on human-generated dataset}  
Table~\ref{tab:results-human} presents the testing results on the human-generated datasets. For example, the rerankers consistently outperform others, showing their reasoning and context understanding ability. Compared to the synthesized datasets, the performance on human-generated queries is generally higher for most models, but the overall trends remain consistent. This indicates that synthesized datasets may be more challenging, highlighting the complexity of the tasks on our synthesized queries. 

Another interesting observation is that the performance of the rerankers is particularly strong on human-generated queries, which may contain more nuanced language and diverse expressions. This suggests that rerankers excel in interpreting and leveraging the richness of human language to improve retrieval accuracy.

\begin{table}[t]
    \centering
    \caption{Latency (s) of the retrieval systems on \benchmarkt. }
    \resizebox{0.9\textwidth}{!}{
    \begin{tabular}{l|cccccc}
        \toprule
        &\textbf{\small DPR} & \textbf{\small QAGNN } & 
        \textbf{\small ada-002} & \textbf{\small multi-ada-002} & \textbf{\small Claude3 Reranker} & \textbf{\small GPT4 Reranker} \\
        \midrule
        \textbf{\amazont} & 2.34 & 2.32 &5.71 &4.87 & 27.24&24.76 \\
        \textbf{\magt} & 0.94 &1.35 &2.25 &3.14 &22.60 & 23.43 \\
        \textbf{\primekgt} & 0.92 & 1.29 &0.54 &0.90 &29.14 &26.97 \\
        \midrule 
        \textbf{Average} & 1.40  & 1.65 &2.83 &2.97 &26.33 &25.05 \\
        \bottomrule
    \end{tabular}
    }
    \label{tab:latency}
    \vspace{-10pt}
\end{table}


\begin{figure}[t]
    \centering
    \includegraphics[width=1\textwidth]{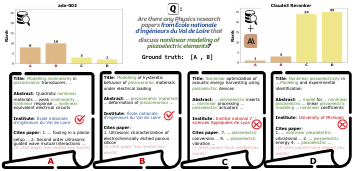}
    \caption{
    A case study on \magh shows that ada-002 overranks non-ground truth papers $C$ and $D$ due to repeated keywords in the relational information ``cites paper''. After reranking with Claude3, it correctly prioritizes ground truth papers $A$ and $B$ with accurate reasoning and analysis.}
    \vspace{-10pt}
    \label{fig:case}
\end{figure}

\xhdr{Retrieval latency} Latency is crucial for practical retrieval systems, as users expect quick responses. As shown in Table~\ref{tab:latency}, we evaluated the latency of various models using a single NVIDIA A100-SXM4-80GB GPU. We observed that the DPR and QAGNN models exhibit lower average latency, making them suitable for time-sensitive applications. In contrast, the ada-002 and multi-ada-002 models have moderate latency due to multiple API calls. However, when combined with LLM rerankers, the latency increases significantly due to the computational demands of these large models. Therefore, it is important to balance accuracy and latency, especially for complex queries that require advanced reasoning capabilities.

\xhdr{Case study} 
To highlight the importance of reasoning ability for achieving good performance on our benchmark, we present a case study in Figure~\ref{fig:case}, comparing the ada-002 model with the Claude3 Reranker. In this example, the query requests papers from a specific institution on a particular topic. The ada-002 model fails to address the relational aspect of the query because it embeds entire documents without detailed analysis. This leads to high relevance scores for irrelevant papers that frequently mention keywords like "nonlinear modeling" and "piezoelectric elements" but do not satisfy the relational requirement. In contrast, the LLM reranker significantly improves the results by reasoning about the relationship between the query and each paper, resulting in scores that more accurately reflect relevance. This underscores the need for reasoning ability to grasp query complexities.

\section{Related Work}
\label{sec:related}

\xhdr{Unstructured QA Datasets}
This research domain consists of methods for retrieving answers from unstructured text, either from a single document~\cite{rajpurkar-etal-2016-squad} or multiple documents~\cite{trischler-etal-2017-newsqa, yang-etal-2018-hotpotqa, TriviaQA, WelblSR18, dunn2017searchqa}. For instance, SQuAD~\cite{rajpurkar-etal-2016-squad} is designed for answer extraction within a specific document, while approaches like HotpotQA~\cite{yang-etal-2018-hotpotqa} and TriviaQA~\cite{TriviaQA} extend to multi-document contexts. Additionally, some studies utilize search engine outputs as a basis or supplementary data for question answering~\cite{DBLP:conf/nips/NguyenRSGTMD16, kwiatkowski-etal-2019-natural}.
However, unstructured QA datasets often lack the depth of relational reasoning commonly required in answering complex user queries. \revision{In contrast, \benchmarkh contains queries demanding multi-hop relational reasoning to challenge model's ability of handling structured information. }
\xhdr{Structured QA Datasets} These datasets challenge models to derive answers from structured sources such as knowledge graphs~\cite{bordes-etal-2014-question,yih-etal-2016-value,talmor-berant-2018-web,journals/corr/BordesUCW15,cao-etal-2022-kqa,GAO_FODOR_KIFER_2019,he2024gretriever} or tabular data~\cite{zhongSeq2SQL2017, yu-etal-2018-spider}. ComplexWebQuestions~\cite{talmor-berant-2018-web} and GraphQA~\cite{he2024gretriever} propose challenges in interpreting complex queries and textualizing graph structures in KBQA, respectively. For tabular data, WikiSQL~\cite{zhongSeq2SQL2017} focuses on translating queries to SQL for single-table databases, whereas Spider~\cite{yu-etal-2018-spider} tackles multi-table scenarios.
\revision{Despite the emphasis on relational data, the restriction to predefined entities and relationships limits the scope of queries. \benchmarkh integrates textual content within structured frameworks to enhance the depth and breadth of information retrieval, promoting richer and more nuanced understanding from extensive textual data.}


\xhdr{Semi-Structured QA Datasets}
This category merges tabular and textual data, presenting challenges in semi-structured data comprehension. WikiTableQuestions~\cite{pasupat-liang-2015-compositional} stresses the integration of table structures with textual elements. TabFact~\cite{Chen2020TabFact:}, HybridQA~\cite{chen-etal-2020-hybridqa}, and TabMCQ~\cite{journals/corr/JauharTH16} extend this by combining validation of textual statements with tabular reasoning. However, datasets leveraging tables as structured frameworks often lack in depicting the rich relational dynamics among entities. Moreover, prior efforts to link textual and tabular information via external sources have led to cumbersome data constructs. \revision{Addressing these challenges, \benchmarkh enhances integration, allowing for flexible navigation and advanced retrieval within complex semi-structured knowledge bases, and facilitating more effective relational reasoning and text handling.}

\vspace{-3pt}
\section{Conclusion and Future Work}
\label{sec:conclusion}
\vspace{-5pt}
We introduce \benchmarkt, the first benchmark to thoroughly evaluate LLM-driven retrieval systems for semi-structured knowledge bases (SKBs). Featuring diverse, natural-sounding queries that require context-specific reasoning across diverse domains, \benchmarkh sets a new standard for assessing real-world retrieval systems. We contribute three large-scale retrieval datasets with human-generated queries and an automated pipeline to simulate realistic user queries. Our experiments on \benchmarkh highlight significant challenges for current models in effectively handling textual and relational information. \benchmarkh paves the way for future research to advance complex, multimodal retrieval systems, focusing on reducing retrieval latency and enhancing reasoning abiliites.

Our current SKBs are limited to textual and relational information. Future work should incorporate additional modalities such as images, videos, and speech to provide a more comprehensive information retrieval system. Despite our anonymization efforts, we acknowledge that privacy remains a potential concern when extending this work to other domains with real user data, which should be protected to ensure compliance with privacy regulations.

\newpage

\section{Acknowledgement}
We thank group members in Leskovec and Zou labs for providing valuable suggestions and conducting benchmark construction. We express our gratitude to the following individuals for their assistance in generating the human-generated queries (ordered by last name):  

Michael Bereket, Charlotte Bunne, Yiqun Chen, Ian Covert, 
Alejandro Dobles, Teddy Ganea, Bryan He, Mika Sarkin Jain, Weixin Liang, Gavin Li, Jiayi Li, 
Sheng Liu, Michael Moor, Hamed Nilforoshan, Rishi Puri, Rishabh Ranjan, Yanay Rosen, Yangyi Shen, Jake Silberg, Elana Simon, Rok Sosic, Kyle Swanson, Nitya Thakkar, Rahul Thapa, Kevin Wu, Eric Wu, Kailas Vodrahalli. 

We especially thank Gavin Li at Stanford University and Zhanghan Wang at New York University for helping build the interactive interface for our SKBs.

We gratefully acknowledge the support of
DARPA under Nos. N660011924033 (MCS);
NSF under Nos. OAC-1835598 (CINES), CCF-1918940 (Expeditions), DMS-2327709 (IHBEM);
Stanford Data Applications Initiative,
Wu Tsai Neurosciences Institute,
Stanford Institute for Human-Centered AI,
Chan Zuckerberg Initiative,
Amazon, Genentech, GSK, Hitachi, SAP, and UCB.

\renewcommand\refname{REFERENCES}
\bibliographystyle{ACM-Reference-Format}
\bibliography{ref}

\section*{Checklist}


\begin{enumerate}

\item For all authors...
\begin{enumerate}
  \item Do the main claims made in the abstract and introduction accurately reflect the paper's contributions and scope?
    \answerYes{} We clearly state our problem scope and highlight our main contribution compared to the existing works.
  \item Did you describe the limitations of your work?
    \answerYes{} Please see the conclusion and future work section.
  \item Did you discuss any potential negative societal impacts of your work?
    \answerYes{} Please see the conclusion and future work section.
  \item Have you read the ethics review guidelines and ensured that your paper conforms to them?
    \answerYes{}
\end{enumerate}

\item If you are including theoretical results...
\begin{enumerate}
  \item Did you state the full set of assumptions of all theoretical results?
    \answerNA{}
	\item Did you include complete proofs of all theoretical results?
    \answerNA{}
\end{enumerate}

\item If you ran experiments (e.g. for benchmarks)...
\begin{enumerate}
  \item Did you include the code, data, and instructions needed to reproduce the main experimental results (either in the supplemental material or as a URL)?
    \answerYes{} Our website {\textcolor{cyan}{\url{https://stark.stanford.edu/}}} and GitHub codebase {\textcolor{cyan}{\url{https://github.com/snap-stanford/STaRK}}} contains code, data, and experimental pipelines.
  \item Did you specify all the training details (e.g., data splits, hyperparameters, how they were chosen)?
    \answerYes{} Please see the experiment setup in Section 3.1, where we explained all of the choices made. We also make the data splits public.
	\item Did you report error bars (e.g., with respect to the random seed after running experiments multiple times)?
    \answerNA{}
	\item Did you include the total amount of compute and the type of resources used (e.g., type of GPUs, internal cluster, or cloud provider)?
    \answerYes{} We provide the GPU device information and report the latency cost.
\end{enumerate}

\item If you are using existing assets (e.g., code, data, models) or curating/releasing new assets...
\begin{enumerate}
  \item If your work uses existing assets, did you cite the creators?
    \answerYes{}
  \item Did you mention the license of the assets?
    \answerYes{} We mentioned them in our webpage.
  \item Did you include any new assets either in the supplemental material or as a URL?
    \answerYes{}
  \item Did you discuss whether and how consent was obtained from people whose data you're using/curating?
    \answerYes{} The resources are from exisiting public data that is open to access.
  \item Did you discuss whether the data you are using/curating contains personally identifiable information or offensive content?
    \answerYes{} The only data with potential personally identifiable information and offensive content is Amazon semi-structured dataset, which is already made anonymized by the public resources.
\end{enumerate}

\item If you used crowdsourcing or conducted research with human subjects...
\begin{enumerate}
  \item Did you include the full text of instructions given to participants and screenshots, if applicable?
    \answerYes{}
  \item Did you describe any potential participant risks, with links to Institutional Review Board (IRB) approvals, if applicable?
    \answerNA{}
  \item Did you include the estimated hourly wage paid to participants and the total amount spent on participant compensation?
    \answerNA{}. We invited volunteered participants who are acknowledged.
    \end{enumerate}

\end{enumerate}

\clearpage

\appendix
\section{Benchmark details}
\label{app:details}

\subsection{Semi-structured Knowledge Bases (SKBs)}
\label{app:skb}

We present the public sources that we used to construct the SKBs in the table below. We have adhered to the licenses of each public resource.

\begin{table}[h]
    \centering
    \caption{Sources of relational structure and textual information of the benchmarks}
    \resizebox{1.0\textwidth}{!}{
    \begin{tabular}{l|cc}
        \toprule
        & \textbf{relational structure} & \textbf{textual information}  \\
        \midrule
        \multirow{ 2}{*}{\amazont} &  \multirow{ 2}{*}{Amazon Product Reviews} & Amazon Product Reviews\\
        & & Amazon Question and Answer Data  \\
        \magt & ogbn-mag & 	ogbn-papers100M, Microsoft Academic Graph  \\
        \multirow{ 2}{*}{\primekgt}& \multirow{ 2}{*}{PrimeKG}  & \texttt{disease}: Orphanet; \texttt{drug}: DrugBank; \texttt{pathway}: Reactome; \\
        && \texttt{gene}: Ensembl, NCBI Entrez, Uniprot, UCSC, CPDB \\
        \bottomrule
    \end{tabular}
    }
    \label{tab:sources}
\end{table}

We build an interactive platform to inspect the data of all three SKBs at \textcolor{cyan}{\url{https://stark.stanford.edu/skb_explorer.html}}. 
We introduce more detailed data statistics below. 

\xhdr{Amazon SKB} In total, it comprises around 1.0M entities (\texttt{product} entities: 0.9M, \texttt{brand} entities: 0.1M, \texttt{category} entities: 1.4k, \texttt{color} entities: 1.7k) and 9.4M relations (\texttt{also\_bought}: 2.8M, \texttt{also\_viewed}: 1.9M, \texttt{has\_brand}: 1.7M, \texttt{has\_category}: 2.3M, \texttt{has\_color}: 0.6M).

\xhdr{MAG SKB}
    This SKB contains around 1.9M entities under four entity types  (\texttt{author}: 1.1M, \texttt{paper}: 0.7M, \texttt{institution}: 8.7K, \texttt{field\_of\_study}: 59.5k) and 39.8M relations under four relation types  (\texttt{author\_writes\_paper}: 13.5M, \texttt{paper\_has\_field\_of\_study}: 14.5M, \texttt{paper\_cites\_paper}: 9.7M, \texttt{author\_affiliated\_with\_institution}: 2.0M). 
    
\xhdr{Prime SKB}
The entity count in our knowledge base is approximately 129.3K, with around 8.1M relations. The numbers of entities in each type are listed below:
\begin{center}
    \begin{verbatim}
    #disease:            17,080 
    #gene/protein:       27,671 
    #molecular_function: 11,169 
    #drug:                7,957 
    #pathway:             2,516 
    #anatomy:            14,035 
    #effect/phenotype:   15,311 
    #biological_process: 28,642 
    #cellular_component:  4,176 
    #exposure:              818
    \end{verbatim}
\end{center}

\subsection{\revision{\amazont}}
\label{app:amazon}


\xhdr{Relational query templates} These are the \textbf{basic} relational templates on \amazont. Note that the final relational template can be composed of multiple basic templates. For example, `\texttt{(color → product ← brand)}' represents a relational template combined from two basic relational templates.
\begin{table}[h]
\begin{tabularx}{\textwidth}{c|X}
\toprule
metapath          & Query template\\
\midrule
\texttt{(brand → product)} & "Can you list the products made by <brand>?" \\
\texttt{(product → product)} & "Which products are similar to <product>?" \\
\texttt{(color → product)} & "Can you provide a list of products that are available in <color>?" \\
\texttt{(category → product)}& "What products are available in the <category> category?" \\
\bottomrule
\end{tabularx}
\end{table}

\subsection{\magt}
\label{app:mag}

\xhdr{Relational query templates} We constructed seven relational templates below:

\begin{table}[h]
\begin{tabularx}{\textwidth}{c|X}
\toprule
metapath          & multi-hop query template\\
\midrule
 \texttt{(author → paper)} & "Can you list the papers authored by <author>?" \\
\texttt{(paper → paper)} &  "Which papers have been cited by the paper <paper>?"\\
\texttt{(field\_of\_study → paper)} & "Can you provide a list of papers in the field of <field\_of\_study>?"\\
\texttt{(institution → author → paper)}& "What papers have been published by researchers from <institution>?"\\
\texttt{(paper → author → paper)}& "What papers have been published by researchers that are coauthors of <paper>?"\\
\makecell{\texttt{(paper → author → paper} \\ \texttt{ ← field\_of\_study ← paper)}}&"Can you find papers that share a coauthor with <paper> and are also in the same field of study?" \\
\makecell{\texttt{(institution → author → } \\ \texttt{paper ← field\_of\_study)}} & "Are there any papers associated with <institution> and are in the field of <field\_of\_study>?" \\
\bottomrule
\end{tabularx}
\end{table}

For example, the metapath \texttt{(field\_of\_study $\rightarrow$ paper)} requires an initial \texttt{field\_of\_study} entity to be filled in the corresponding query template. For multi-hop metapaths, the last metapath \texttt{(institution $\rightarrow$ author $\rightarrow$ paper $\leftarrow$ field\_of\_study)} requires an \texttt{institution} entity and a \texttt{field\_of\_study} entity to initialize the query. 

\subsection{\primekgt}
\label{app:prime}

\xhdr{Relational query templates} For synthesized queries, we listed 28 multi-hop templates designed by experts to cover various relation types and ensure their practical relevance. 

For instance, the query ``What is the drug that targets the genes or proteins expressed in <anatomy>?'' serves applications in precision medicine and pharmacogenomics, 
aiding researchers and healthcare professionals in identifying drugs that act on genes or proteins associated with specific anatomical areas and enabling more targeted treatments. 

\vspace{10pt}
\begin{mdframed}[backgroundcolor=gray!5, linecolor=black, linewidth=2pt, innerleftmargin=10pt, innerrightmargin=10pt, innertopmargin=3pt, innerbottommargin=10pt]
\begin{scriptsize}
\begin{verbatim}
{   
(effect/phenotype → [phenotype absent] → disease ← [!indication] ← drug): 
    "Find diseases with zero indication drug and are associated with <effect/phenotype>",
(drug → [contraindication] → disease ← [associated with] ← gene/protein): 
    "Identify diseases associated with <gene/protein> and are contraindicated with <drug>",
(anatomy → [expression present] → gene/protein ← [expression absent] ← anatomy): 
    "What gene or protein is expressed in <anatomy1> while is absent in <anatomy2>?",
(anatomy → [expression absent] → gene/protein ← [expression absent] ← anatomy): 
    "What gene/protein is absent in both <anatomy1> and <anatomy2>?",
(drug → [carrier] → gene/protein ← [carrier] ← drug): 
    "Which target genes are shared carriers between <drug1> and <drug2>?",
(anatomy → [expression present] → gene/protein → [target] → drug): 
    "What is the drug that targets the genes or proteins which are expressed in <anatomy>?",
(drug → [side effect] → effect/phenotype → [side effect] → drug): 
    "What drug has common side effects as <drug>?",
(drug → [carrier] → gene/protein → [carrier] → drug): 
    "What is the drug that has common gene/protein carrier with <drug>?",
(anatomy → [expression present] → gene/protein → enzyme → drug): 
    "What is the drug that some genes or proteins act as an enzyme upon, 
     where the genes or proteins are expressed in <anatomy>?",
(cellular_component → [interacts with] → gene/protein → [carrier] → drug): 
    "What is the drug carried by genes or proteins that interact with <cellular_component>?",
(molecular_function → [interacts with] → gene/protein → [target] → drug): 
    "What drug targets the genes or proteins that interact with <molecular_function>?",
(effect/phenotype → [side effect] → drug → [synergistic interaction] → drug): 
    "What drug has a synergistic interaction with the drug that has <effect/phenotype> 
    as a side effect?",
(disease → [indication] → drug → [contraindication] → disease): 
    "What disease is a contraindication for the drugs indicated for <disease>?",
(disease → [parent-child] → disease → [phenotype present] → effect/phenotype): 
    "What effect or phenotype is present in the sub type of <disease>?",
(gene/protein → [transporter] → drug → [side effect] → effect/phenotype): 
    "What effect or phenotype is a [side effect] of the drug transported by <gene/protein>?",
(drug → [transporter] → gene/protein → [interacts with] → exposure): 
    "What exposure may affect <drug>s efficacy by acting on its transporter genes?",
(pathway → [interacts with] → gene/protein → [ppi] → gene/protein): 
    "What gene/protein interacts with the gene/protein that related to <pathway>?",
(drug → [synergistic interaction] → drug → [transporter] → gene/protein): 
    "What gene or protein transports the drugs that have a synergistic interaction with <drug>?",
(biological_process → [interacts with] → gene/protein → [interacts with] → biological_process): 
    "What biological process has the common interactino pattern with gene or proteins as 
    <biological_process>?",
(effect/phenotype → [associated with] → gene/protein → [interacts with] → biological_process): 
    "What biological process interacts with the gene/protein associated with <effect/phenotype>?",
(drug → [transporter] → gene/protein → [expression present] → anatomy): 
    "What anatomy expressesed by the gene/protein that affect the transporter of <drug>?",
(drug → [target] → gene/protein → [interacts with] → cellular_component): 
    "What cellular component interacts with genes or proteins targeted by <drug>?",
(biological_process → [interacts with] → gene/protein → [expression absent] → anatomy): 
    "What anatomy does not express the genes or proteins that interacts with <biological_process>?",
(effect/phenotype → [associated with] → gene/protein → [expression absent] → anatomy): 
    "What anatomy does not express the genes or proteins associated with <effect/phenotype>?",
(drug → [indication] → disease → [indication] → drug) 
 & (drug → [synergistic interaction] → drug): 
    "Find drugs that has a synergistic interaction with <drug> and both are indicated 
    for the same disease.",
(pathway → [interacts with] → gene/protein → [interacts with] → pathway) 
    & (pathway → [parent-child] → pathway): 
    "Find pathway that is related with <pathway> and both can [interacts with] the same gene/protein.",
(gene/protein → [associated with] → disease → [associated with] → gene/protein) 
    & (gene/protein → [ppi] → gene/protein): 
    "Find gene/protein that can interect with <gene/protein> and both are associated 
    with the same disease.",
(gene/protein → [associated with] → effect/phenotype → [associated with] → gene/protein) 
    & (gene/protein → [ppi] → gene/protein): 
    "Find gene/protein that can interect with <gene/protein> and both are associated 
    with the same effect/phenotype."
}
\end{verbatim}
\end{scriptsize}
\end{mdframed}

where $[\cdot]$ denotes the relation type.
\section{\revision{Mathematical Definitions of Shannon Entropy and Type-Token Ratio}}
\label{app:math}
\textbf{Shannon Entropy.} Shannon Entropy is a measure of the uncertainty in a set of possible outcomes, quantifying the amount of information or disorder within a dataset. It is defined as follows:

\[
H(X) = - \sum_{i=1}^{n} p(x_i) \log p(x_i)
\]

where \( X \) is the set of possible outcomes, \( p(x_i) \) is the probability of occurrence of the outcome \( x_i \), and \( n \) is the total number of unique outcomes. Higher entropy values indicate greater diversity in the distribution of outcomes.\cite{https://doi.org/10.1002/j.1538-7305.1948.tb01338.x}

\textbf{Type-Token Ratio (TTR).} The Type-Token Ratio is a measure of lexical diversity, calculated as the ratio of the number of unique words (types) to the total number of words (tokens) in a text. It is defined as follows:

\[
\text{TTR} = \frac{V}{N}
\]

where \( V \) is the number of unique words and \( N \) is the total number of words in the text. Higher TTR values indicate a higher proportion of unique words, reflecting greater lexical diversity. \cite{262c3ab1-41bb-3b12-a85c-0620dcf5ed8c}

\section{Instructions for Generating Queries}
\label{app:instruction}

For the process of generating queries by human, the participants were given a list of entity IDs that we randomly sampled from the entire entity set. Then, they were asked to follow the following instructions with the support of our built interactive platform at {\textcolor{cyan}{\url{https://stark.stanford.edu/skb_explorer.html}}}. 

\vspace{10pt}
\begin{mdframed}[backgroundcolor=gray!10, linecolor=black, linewidth=2pt, innerleftmargin=10pt, innerrightmargin=10pt, innertopmargin=3pt, innerbottommargin=10pt]
\textbf{Task: }

1) Given the provided entity ID, review the associated document and any connected entities and multi-hop paths.

2) Find interesting aspects of the entities by examining both their relational structures and the textual information available.

3) Write your queries from these aspects such that the entity can satisfy all of them. 

\textbf{Note:}

1) Please do not leak the name of the entity in the query. 

2) You can skip some entity IDs if you think the knowledge involved is hard to understand. 

3) Feel free to be creative with content of your queries, you can also include additional context. There is NO restriction on how you express the queries.
\end{mdframed}
After collecting the queries, we filtering the ground truth answers manually by human validation. 

\section{Experiments}




\subsection{More Experimental Results}
\label{app:exp}

\begin{table}[h]
    \centering
    \caption{Positive/Non-negative rates (\%) from human evaluation.}
    \begin{tabular}{l|ccc}
        \toprule
        & Naturalness & Diversity & Practicality \\
        \midrule
        \amazonh &  73.6 / 89.5 & 68.4 / 89.5  & 89.5 / 94.7\\
        \magh & 94.7 / 100 & 73.7 / 84.2 & 68.4 / 84.2 \\
        \primekgh & 67.8 / 92.8 & 71.4 / 82.1 & 71.4 / 89.3\\
        \midrule
        Average & 78.7 / 94.1 & 71.0 / 85.3 & 76.4 / 89.4\\
        \bottomrule
    \end{tabular}
    \label{tab:human}
\end{table}

\section{Prompts and LLM versions for Query Synthesization}
\label{app:prompts}

We summarize the LLM versions in Table~\ref{tab:llm_versions_datasets}. 
We chose these models based on a joint consideration of their cost, how accurate they are, and whether they were the latest model during different phases of the project. While we used different LLMs, we checked each step separately to make sure the good quality in our benchmark datasets.

\begin{table}[h]
    \centering
    \caption{Summary of LLM Versions for Query Synthesization}
    \resizebox{\textwidth}{!}{ 
    \begin{tabular}{l|ccc}
        \toprule
        \textbf{Step} & \amazonh & \magh & \primekgh \\
        \midrule
        Step 2: Extracting textual requirements & \texttt{gpt-3.5-turbo-16k} & \texttt{claude-2.0} & \texttt{claude-2.0} \\
        \midrule
        Step 3: Combining relational and textual requirements & \multicolumn{3}{c}{\texttt{claude-2.0}, \texttt{gpt-4-0125-preview}} \\
        \midrule
        Step 4: Filtering additional answers & \multicolumn{3}{c}{\texttt{claude-2.1}, \texttt{claude-2.0}, \texttt{claude-instant-1.2}} \\
        \bottomrule
    \end{tabular}
    }
    \label{tab:llm_versions_datasets}
\end{table}

\subsection{Extracting textual requirements}
\label{app:text}




\begin{mdframed}[backgroundcolor=cyan!5, linecolor=black, linewidth=2pt, 
                innerleftmargin=10pt, innerrightmargin=10pt, 
                innertopmargin=3pt, innerbottommargin=10pt]
\section*{Prompt for \amazont: Textual requirement extraction}
\begin{lstlisting}
You are an intelligent assistant that extracts diverse positive requirements and negative perspectives for an Amazon product. I will give you the following information:
- product: <product name>
- dimensions: <product dimensions>
- weight: <product weight>
- description: <product description>
- features: #1: <feature #1>  ...
- reviews: 
  #1:
    summary: <review summary>
    text: <full review text>
  #2: ...
- Q&A: 
  #1:
    question: <product-related question>
    answer: <answer to product-related question>
  #2: ...
Based on the given product information, you need to (1) identify the product's generic category, (2) list all of the negative perspectives and their sources, and (2) extract up to five hard and five soft requirements relevant to customers' interests along with their sources. (1) For example, the product's generic category can be "a chess book" or "a phone case for iphone 6", do not use the product name directly. (2) Negative perspectives are those that the product doesn't fulfill, which come from the negative reviews or Q&A. (3) For the requirements, you should only focus on the product's advantages and positive perspects. Hard requirements mean that product must fulfil, such as size and functionality. Soft requirements are not as strictly defined but still desirable, such as a product is easy-to-use. For (2) and (3), each source is a composite of the key and index (if applicable) separated by "-", such as "description", "Q&A-#1". You should provide the response in a specific format as follows where "item" refers to the product's generic category, e.g., "a chess book". 
Response format:
{
  "item": <the product's generic category> ,
  "negative": [[<source of negative perspective>, <negative perspective description>]],
  "hard": [[<source of hard requirement>, <hard requirement description>], ...],
  "soft": [[<source of soft requirement>, <soft requirement description>], ...]
}
Here is an example of the response: 
{
  "item": "a camping chair",
  "negative": [["reviews-#3", "the chair is not sturdy enough"], ["Q&A-#1", "wrong color"]],
  "hard": [["description", "has a breathable mesh back"], ["description", "the arm is adjustable"], ["dimensions", "more than 35 inches long"], ["features-#7", "with a arm rest cup holder"], ["Q&A-#4", "need to come with a carrying bag"]],
  "soft": [["description", "suitable for outdoors"], ["features-#9", "compact and save space"], ["reviews-#6", "light and portable"]]
}
This is the information of the product that you need to write response for:
<product_doc>
Response: 
\end{lstlisting}
\end{mdframed}

\begin{mdframed}[backgroundcolor=pink!25, linecolor=black, linewidth=2pt, innerleftmargin=10pt, innerrightmargin=10pt, innertopmargin=3pt, innerbottommargin=10pt]
\section*{Prompt for \magt: Textual requirement extraction}
\begin{lstlisting}
You are a helpful assistant that helps me extract one short requirement (no more than 10 words) about a paper from the paper information that researchers might be interested in. The requirement can be about the paper content, publication date, publication venue, etc. The requirement should be general and not too specific. I will give you the paper information, and you should return a short phrase about the paper, starting with 'the paper...'. This is the paper information:
<doc_info>
Please only return the short and general requirement without additional comments.
\end{lstlisting}
\end{mdframed}

\begin{mdframed}[backgroundcolor=lime!10, linecolor=black, linewidth=2pt, innerleftmargin=10pt, innerrightmargin=10pt, innertopmargin=3pt, innerbottommargin=10pt]
\section*{Prompt for \primekgt: Textual requirement extraction}
\begin{lstlisting}
You are a helpful assistant that helps me extract <n_properties> from a given <target> information that a <role> may be interested in. 
<role_instruction> 
Each property should be no more than 10 words and start with "the <target>". You should also include the source of each property as indicated in the paragraph names of the information, e.g., "details.mayo_symptoms", "details.summary", etc. You should return a list of properties and their sources following the format:
[["<short_property1>", "<source1>"], ["<short_property2>", "<source2>"], ...]
This is the information:
<doc_info>
Please provide only the list with <n_properties> in your response. Response:
\end{lstlisting}
\end{mdframed}

According to the role assigned to simulate the query content, the \texttt{<role\_instruction>} as shown below is filled in accordingly.
\begin{table}[h]
\centering
\begin{tabularx}{\textwidth}{lX}
\toprule
role          & role instruction\\
\midrule
 Doctor & {Doctors typically ask questions aimed at diagnosing and treating. Their questions tend to be direct and practical, focusing on aspects involving side effects, symptoms, and complications etc.}\\
 \noalign{\vspace{1.5ex}}
 Medical scientist & {Medical scientists often ask questions that reflect the complexity and depth of the scientific inquiry in the medical field. Their questions tend to be detailed and specific, focusing on aspects such as: etiology and pathophysiology, genetic factors, association with pathway, protein, or molecular function. }\\
 \noalign{\vspace{1.5ex}}
 Patient & {Patients typically don't know the professional medical terminology. Their questions tend to be straightforward, focusing on practical concerns on the symptons, effects, and inheritance etc., instead of the detailed mechanisms, which may also include more context. }\\
\bottomrule
\end{tabularx}
\end{table}

\subsection{Combining relational and textual requirements}
\label{app:fusion}

\begin{mdframed}[backgroundcolor=cyan!5, linecolor=black, linewidth=2pt, innerleftmargin=10pt, innerrightmargin=10pt, innertopmargin=3pt, innerbottommargin=10pt]
\section*{Prompt for \amazont: Fuse relational and textual requirements}
\begin{lstlisting}
You are an intelligent assistant that generates queries about an Amazon item. I will provide you with the item name, requirements, and its negative customer reviews. Your task is to create a natural-sounding customer query that leads to the item as the answer, using the requirements that are non-conflicting with the negative reviews, and provide the indices of the requirements used. For example:

Information:
- item: a soccer rebounder
- requirements:
#1: needs a heavy-duty 1-inch to 3-inch steel tube frame
#2: should be adjustable for practicing different skills
#3: should be durable
#4: usually be viewed together with <SKLZ Star-Kick Hands Free Solo Soccer Trainer>
- negative reviews:
#1: it was broken after a few uses

Response:
{
  "index": [1, 2, 4],
  "query": "Please recommend a soccer rebounder with a steel frame, about 2 inches thick, that can adapt to different skill levels. We had a blast using the <SKLZ Star-Kick Hands Free Solo Soccer Trainer> with my family, and I'm on the lookout for something similar." 
}

As the negative review indicates that the soccer rebouncer lacks durability, your query should only incorporate requirements #1, #2, and #4 while excluding #3. A requirement should only be excluded if it conflicts with negative feedback or is unlikely to align with customers' interests. For relational requirements about another <product>, do not directly use "usually bought/viewed together with <product>" in the query. You must deduce the item's relationship with <product> into substitute or complement, and create various user scenarios, such as the item should be compatible or used with <product> (for complements) or match in style with <product> (for substitute), to make the queries sound natural. Except for <product>, you should change the description but convey similar meanings. The query structure is completely flexible. Here is the information to generate the requirement indices and a natural-sounding query:

Information:
<product_req_and_neg_comments>
Response:
\end{lstlisting}
\end{mdframed}

\begin{mdframed}[backgroundcolor=lime!10, linecolor=black, linewidth=2pt, innerleftmargin=10pt, innerrightmargin=10pt, innertopmargin=3pt, innerbottommargin=10pt]
\section*{Prompt for \magt: Fuse relational and textual requirements}
\begin{lstlisting}
You are a helpful assistant that helps me generate a new query by incorporating an additional requirement into a given query, and form a coherent and natural-sounding question. 
This is the existing query: 
<query>
This is the additional requirement: 
<additional_textual_requirement>
You should be creative in combining the existing query and requirement, and flexible in structuring the new query, adding context as needed. Please return the new query without additional comments:
\end{lstlisting}

The prompt of a second-time rewrite by GPT-4 Turbo:
\begin{lstlisting}
You are a helpful assistant that helps make a researcher's query about a paper more natural-sounding, akin to the language used in ArXiv web searches. You should change the description but convey similar meanings. The query structure is completely flexible. The original query:
"<query>"
Please only output the new query without additional comments:
\end{lstlisting}
\end{mdframed}

\begin{mdframed}[backgroundcolor=pink!25, linecolor=black, linewidth=2pt, innerleftmargin=10pt, innerrightmargin=10pt, innertopmargin=3pt, innerbottommargin=10pt]
\section*{Prompt for \primekgt: Fuse relational and textual requirements}
\begin{lstlisting}
You are a helpful assistant that helps me generate a natural-sounding and coherent query as if you were a <role>. The query should be created based on a list of requirements for searching <plural_target> in a database. I will provide you with the requirements in the following format:
[<requirement1>, <requirement2>, ...]
You should create the query based solely on the given requirements. Moreover, you should craft the query from the perspective of a <role>. 
<role_instruction>
For example, a query from a <role> could be
"<example_query>"
You can be flexible in structuring the query and adding additional context. Ensure that the query uses different descriptions than the original property descriptions while retaining similar meanings. The query should sound concise and natural. These are the requirements:
<requirements>
Please create the query based on the given requirements and provide only the query without additional comments. Your response:
\end{lstlisting}

The prompt of a second-time rewrite by GPT-4 Turbo:
\begin{lstlisting}
You are a helpful assistant that helps me rewrite a query that searches for <plural_target> from the perspective of a <role>. You should maintain the requirements from the original query and the characteristics of the <role>, while being creative and flexible in structuring the query. Ensure the revised query is concise and natural-sounding. Original query: "<query>". Please output only the rewritten query:
\end{lstlisting}
\end{mdframed}

\subsection{Filtering additional answers}
\begin{mdframed}[backgroundcolor=cyan!5, linecolor=black, linewidth=2pt, innerleftmargin=10pt, innerrightmargin=10pt, innertopmargin=3pt, innerbottommargin=10pt]
\section*{Prompt for \amazont: Filtering additional answers}
\subsection*{Filter products by general category}
\begin{lstlisting}
You are an intelligent assistant that identifies whether an Amazon product belongs to a given category. I will give you the product information. You should only answer yes / no in the response. For examples, the product <SKLZ Star-Kick Hands Free Solo Soccer Trainer> belongs to the category "soccer trainer" and the product <Test your Opening, Middlegame and Endgame Play - VOLUME 2> belongs to the category "a chess opening book", while <Baby Girls One-piece Shiny Athletic Leotard Ballet Tutu with Bow> doesn't belong to category "an adult tutu".

Information:
- product title: <<product_title>>
- product description: <product_description>

Does the product belong to "<target_category>"? Response (yes/no): 
\end{lstlisting}

\subsection*{Filter products by requirements}
\begin{lstlisting}
You are a helpful assistant that helps me check whether an Amazon product satisfies the given requirements. I will provide you with the product information, which may include the product description, features, reviews, and Q&A from customers. Your task is to assess whether the product meets each requirement based on the provided information. If there is no information that supports the requirement, your response for that requirement is "NA". If there is relevant information that supports the requirement, your response for that requirement is the information source that fulfills the requirement. Each information source is a composite of the key and index (if applicable), separated by "-", such as "description", "features-#3", "Q&A-#1", "reviews-#2". If there are multiple sources, 

Response:
{
    1: "NA" or [the information sources that satisfy the requirement #1], 
    2: "NA" or [the information sources that satisfy the requirement #2], 
    ...
}

Here is the product information:
<product_doc>
The requirements are as follows:
<customer_requirements>

Response: 
\end{lstlisting}
\end{mdframed}

\begin{mdframed}[backgroundcolor=lime!10, linecolor=black, linewidth=2pt, innerleftmargin=10pt, innerrightmargin=10pt, innertopmargin=3pt, innerbottommargin=10pt]
\section*{Prompt for \magt: Filtering additional answers}
\begin{lstlisting}
You are a helpful assistant that helps me verify whether a given <target_node_type> is subject to a requirement. I will provide you with the <target_node_type> information and the requirement, and you should return only a 'True' or 'False' value, indicating whether the <target_node_type> meets the requirement.
This is the <target_node_type> information:
<doc_info>
This is the requirement:
<additional_textual_requirement>
Please return only the boolean value without additional comments:
\end{lstlisting}
\end{mdframed}

\begin{mdframed}[backgroundcolor=pink!25, linecolor=black, linewidth=2pt, innerleftmargin=10pt, innerrightmargin=10pt, innertopmargin=3pt, innerbottommargin=10pt]
\section*{Prompt for \primekgt: Filtering additional answers}
\begin{lstlisting}
You are a helpful assistant tasked with verifying whether a given <target> satisfies each of the provided requirements. I will give you the requirements in the following format:
{1: <requirement1>, 2: <requirement2>, ...}
When evidence in the <target> information confirms a requirement is met, cite the source, for example, 'details.mayo_symptoms', 'details.summary'. If no direct evidence exists, indicate this with 'NA'. The output in JSON format should be as follows:
{1: 'NA' or <source1>, 2: 'NA' or <source2>, ...}
This is the <target> information:
<doc_info>
These are the requirements:
<requirements>
Please provide only the JSON in your response. Response:
\end{lstlisting}
\end{mdframed}

\clearpage
\end{document}


\renewcommand{\thefootnote}{\fnsymbol{footnote}}

\maketitle
\vspace{-10pt}


\xhdr{Website/Platform and Hosting} 
\vspace{-5pt}
\begin{itemize}[leftmargin=*]
    \item \textbf{Data downloading}: We host our retrieval dataset and knowledge base data in both Hugging Face dataset repository: \href{https://huggingface.co/datasets/snap-stanford/stark/tree/main}{\textcolor{Violet}{https://huggingface.co/datasets/snap-stanford/stark}} and project website under Stanford Computer Science Server: \href{https://stark.stanford.edu/datasets.html}{\textcolor{Violet}{https://stark.stanford.edu/datasets.html}}. 
    \item \textbf{Code}: We release a PyPI package, \texttt{stark-qa} (\href{https://pypi.org/project/stark-qa/}{\textcolor{Violet}{https://pypi.org/project/stark-qa/}}). This package automatically downloads and loads our query and semi-structured datasets. Additionally, we host our developer code on GitHub at \href{https://github.com/snap-stanford/stark}{\textcolor{Violet}{https://github.com/snap-stanford/stark}}, allowing users to submit issues and pull requests.
    \item \textbf{Interactive platform}: We build a platform, \benchmarkh Semi-structured Knowledge Base (SKB) Explorer, for users to inspect our SKB schema: \href{https://stark.stanford.edu/skb_explorer.html}{\textcolor{Violet}{https://stark.stanford.edu/skb\_explorer.html}}.
\end{itemize}

\xhdr{Croissant Metadata} The Croissant metadata for our dataset is available for viewing and downloading at \href{https://stark.stanford.edu/files/croissant_metadata.json}{\textcolor{Violet}{https://stark.stanford.edu/files/croissant\_metadata.json}}. 

\xhdr{DOI} We provide a persistent dereferenceable identifier DOI: https://doi.org/10.57967/hf/2530.

\xhdr{Licensing} The \benchmarkh retrieval datasets are under license CC-BY-4.0 as stated in our website. And our released code is under MIT license, as stated in the GitHub repository.

\xhdr{Maintenance Plan} We plan to update our website with the most recent document and Python package. We will maintain our GitHub repository will pull requests and open issues.

\xhdr{Author Statement} We hereby confirm that we bear all responsibility for any violation of rights that may occur in the use or distribution of the data and content presented in this work. We affirm that we have obtained all necessary permissions and licenses for the data and content included in this work. We confirm that the use of this data complies with all relevant laws and regulations, and we take full responsibility for addressing any claims or disputes that may arise regarding rights violations or licensing issues. 

\xhdr{Reproducibility} We make the following efforts to ensure reproducibility:
\begin{itemize}[leftmargin=*]
    \item \textbf{Code:} We have provided the complete codebase in our GitHub repository. The code includes scripts for data loading, preprocessing, embedding generation, and evaluation. Detailed instructions for setting up the environment and running the code are included to facilitate easy reproduction of the results.
    \item \textbf{Pre-generated Embeddings:} The original embeddings used in the experiments are available for download from our repository. These embeddings can be used directly to replicate the experiments without the need for re-computation, ensuring consistency in results.
    \item \textbf{Evaluation Procedures:} All evaluation procedures are thoroughly documented. Users can follow the provided scripts and guidelines to perform evaluations. We have included detailed steps and examples both in the GitHub repository and on our website to guide users through the entire process, helping them achieve reproducible results.
\end{itemize}

\xhdr{Datasheet for Dataset (Dataset documentation and intended uses)} 
\small
\begin{mdframed}[linecolor=\sectioncolor]
\subsubsection*{\textcolor{\sectioncolor}{
    MOTIVATION
}}
\end{mdframed}

    \textcolor{\sectioncolor}{\textbf{
    For what purpose was the dataset created?
    }
    Was there a specific task in mind? Was there
    a specific gap that needed to be filled? Please provide a description.
    } \\
    The datasets were created to evaluate retrieval tasks on Semi-structured Knowledge Bases (SKB). They are specifically designed to assess the capabilities of current Large Language Models (LLMs) in performing complex retrieval tasks. This initiative aims to address the gap in evaluating how well LLMs can handle retrieval tasks that involve a combination of structured and unstructured data.\\
    
    \textcolor{\sectioncolor}{\textbf{
    Who created this dataset (e.g., which team, research group) and on behalf
    of which entity (e.g., company, institution, organization)?
    }
    } \\
    The datasets were created by lab members in Leskovec and Zou's groups at Stanford University, in collaboration with scientists from Amazon. \\
    
    \textcolor{\sectioncolor}{\textbf{
    What support was needed to make this dataset?
    }
    (e.g.who funded the creation of the dataset? If there is an associated
    grant, provide the name of the grantor and the grant name and number, or if
    it was supported by a company or government agency, give those details.)
    } \\
    We acknowledge the support of
    DARPA under Nos. N660011924033 (MCS);
    NSF under Nos. OAC-1835598 (CINES), CCF-1918940 (Expeditions), DMS-2327709 (IHBEM);
    Stanford Data Applications Initiative,
    Wu Tsai Neurosciences Institute,
    Stanford Institute for Human-Centered AI,
    Chan Zuckerberg Initiative,
    Amazon, Genentech, GSK, Hitachi, SAP, and UCB. \\
    
    \textcolor{\sectioncolor}{\textbf{
    Any other comments?
    }} \\
    No. \\

\begin{mdframed}[linecolor=\sectioncolor]
\subsubsection*{\textcolor{\sectioncolor}{
    COMPOSITION
}}
\end{mdframed}
    \textcolor{\sectioncolor}{\textbf{
    What do the instances that comprise the dataset represent (e.g., documents,
    photos, people, countries)?
    }
    Are there multiple types of instances (e.g., movies, users, and ratings;
    people and interactions between them; nodes and edges)? Please provide a
    description.
    } \\
    The retrieval dataset represents queries (textual data) from people. The instances in the underlying knowledge base represent documents and interactions (i.e., relational data). For example, in the Amazon domain, instances may include product descriptions (documents) and the relationships between products and their categories, brand, and colors (relational data). This combination of textual and relational data allows for a comprehensive evaluation of retrieval tasks on semi-structured knowledge bases.\\
    
    \textcolor{\sectioncolor}{\textbf{
    How many instances are there in total (of each type, if appropriate)?
    }
    } \\
   Please refer to our main paper and appendix for comprehensive statistics. \\
    
    \textcolor{\sectioncolor}{\textbf{
    Does the dataset contain all possible instances or is it a sample (not
    necessarily random) of instances from a larger set?
    }
    If the dataset is a sample, then what is the larger set? Is the sample
    representative of the larger set (e.g., geographic coverage)? If so, please
    describe how this representativeness was validated/verified. If it is not
    representative of the larger set, please describe why not (e.g., to cover a
    more diverse range of instances, because instances were withheld or
    unavailable).
    } \\
    We have released the complete datasets, containing all queries and available entities in the knowledge bases. \\
    
    \textcolor{\sectioncolor}{\textbf{
    What data does each instance consist of?
    }
    “Raw” data (e.g., unprocessed text or images) or features? In either case,
    please provide a description.
    } \\
    We have provided the raw data, which includes the raw query data and the original textual and relational information. \\
    
    \textcolor{\sectioncolor}{\textbf{
    Is there a label or target associated with each instance?
    }
    If so, please provide a description.
    } \\
    Yes, the answer IDs for each query indicate the ground truth entities or items that satisfy the requirements of the query. \\
    
    \textcolor{\sectioncolor}{\textbf{
    Is any information missing from individual instances?
    }
    If so, please provide a description, explaining why this information is
    missing (e.g., because it was unavailable). This does not include
    intentionally removed information, but might include, e.g., redacted text.
    } \\
    No, we did not omit any information. However, some information in the original public resources might not be complete or up-to-date. \\
    
    \textcolor{\sectioncolor}{\textbf{
    Are relationships between individual instances made explicit (e.g., users’
    movie ratings, social network links)?
    }
    If so, please describe how these relationships are made explicit.
    } \\
    Yes. For retrieval datasets, we assume each query is independent. For knowledge base data, relationships between individual entities are made explicit. These relationships are represented through structured data within the knowledge bases, such as entity links and relational attributes. For example, in the Amazon domain, explicit relationships include the connections between products and their categories, brands, and attributes like color. This structured relational data helps in accurately defining the interactions and associations between different entities within the knowledge base. \\
    
    \textcolor{\sectioncolor}{\textbf{
    Are there recommended data splits (e.g., training, development/validation,
    testing)?
    }
    If so, please provide a description of these splits, explaining the
    rationale behind them.
    } \\
    Yes, we have provided the official split for our synthesized queries. \\
    
    \textcolor{\sectioncolor}{\textbf{
    Are there any errors, sources of noise, or redundancies in the dataset?
    }
    If so, please provide a description.
    } \\
    The ground truth answers for each synthesized query are filtered automatically through Large Language Models (LLMs). Therefore, there is a possibility of including missing answers or false answers in the ground truth. However, we have devoted great efforts to reduce the errors.   \\
    
    \textcolor{\sectioncolor}{\textbf{
    Is the dataset self-contained, or does it link to or otherwise rely on
    external resources (e.g., websites, tweets, other datasets)?
    }
    If it links to or relies on external resources, a) are there guarantees
    that they will exist, and remain constant, over time; b) are there official
    archival versions of the complete dataset (i.e., including the external
    resources as they existed at the time the dataset was created); c) are
    there any restrictions (e.g., licenses, fees) associated with any of the
    external resources that might apply to a future user? Please provide
    descriptions of all external resources and any restrictions associated with
    them, as well as links or other access points, as appropriate.
    } \\
    The datasets are self-contained. \\
    
    \textcolor{\sectioncolor}{\textbf{
    Does the dataset contain data that might be considered confidential (e.g.,
    data that is protected by legal privilege or by doctor-patient
    confidentiality, data that includes the content of individuals’ non-public
    communications)?
    }
    If so, please provide a description.
    } \\
    No, the dataset does not contain any data that might be considered confidential. \\
    
    \textcolor{\sectioncolor}{\textbf{
    Does the dataset contain data that, if viewed directly, might be offensive,
    insulting, threatening, or might otherwise cause anxiety?
    }
    If so, please describe why.
    } \\
    No, as far as we are aware, the dataset does not contain any data that might be offensive, insulting, threatening, or cause anxiety. \\
    
    \textcolor{\sectioncolor}{\textbf{
    Does the dataset relate to people?
    }
    If not, you may skip the remaining questions in this section.
    } \\
    The query dataset does not target individuals; however, human efforts were involved in creating a small set of queries and filtering ground truth answers. \\
    
    \textcolor{\sectioncolor}{\textbf{
    Does the dataset identify any subpopulations (e.g., by age, gender)?
    }
    If so, please describe how these subpopulations are identified and
    provide a description of their respective distributions within the dataset.
    } \\
    Not applicable. \\
    
    \textcolor{\sectioncolor}{\textbf{
    Is it possible to identify individuals (i.e., one or more natural persons),
    either directly or indirectly (i.e., in combination with other data) from
    the dataset?
    }
    If so, please describe how.
    } \\
    Not applicable.  \\
    
    \textcolor{\sectioncolor}{\textbf{
    Does the dataset contain data that might be considered sensitive in any way
    (e.g., data that reveals racial or ethnic origins, sexual orientations,
    religious beliefs, political opinions or union memberships, or locations;
    financial or health data; biometric or genetic data; forms of government
    identification, such as social security numbers; criminal history)?
    }
    If so, please provide a description.
    } \\
    Not applicable. \\
    
    \textcolor{\sectioncolor}{\textbf{
    Any other comments?
    }} \\
    No. \\

\begin{mdframed}[linecolor=\sectioncolor]
\subsubsection*{\textcolor{\sectioncolor}{
    COLLECTION
}}
\end{mdframed}

    \textcolor{\sectioncolor}{\textbf{
    How was the data associated with each instance acquired?
    }
    Was the data directly observable (e.g., raw text, movie ratings),
    reported by subjects (e.g., survey responses), or indirectly
    inferred/derived from other data (e.g., part-of-speech tags, model-based
    guesses for age or language)? If data was reported by subjects or
    indirectly inferred/derived from other data, was the data
    validated/verified? If so, please describe how.
    } \\
    The data associated with each instance was acquired through direct observation of raw text data and structured relational data from publicly available sources. Additionally, human evaluators synthesized and validated queries and ground truth answers to ensure accuracy and relevance.\\
    
    \textcolor{\sectioncolor}{\textbf{
    Over what timeframe was the data collected?
    }
    Does this timeframe match the creation timeframe of the data associated
    with the instances (e.g., recent crawl of old news articles)? If not,
    please describe the timeframe in which the data associated with the
    instances was created. Finally, list when the dataset was first published.
    } \\
    The data was collected over a period of several months in 2023. This timeframe aligns with the creation timeframe of the data associated with the instances. The dataset was first published in Apr 2024.\\
    
    \textcolor{\sectioncolor}{\textbf{
    What mechanisms or procedures were used to collect the data (e.g., hardware
    apparatus or sensor, manual human curation, software program, software
    API)?
    }
    How were these mechanisms or procedures validated?
    } \\
    The data was collected using a combination of software programs and APIs to gather raw text and relational data. Manual human curation was employed to synthesize and validate the data, ensuring its accuracy and relevance. The procedures were validated through rigorous human evaluation and analysis.\\
    
    \textcolor{\sectioncolor}{\textbf{
    What was the resource cost of collecting the data?
    }
    (e.g. what were the required computational resources, and the associated
    financial costs, and energy consumption - estimate the carbon footprint.
    See Strubell \textit{et al.}\cite{strubellEnergyPolicyConsiderations2019} for approaches in this area.)
    } \\
    The resource cost of collecting the data included computational resources for running software programs and APIs. API costs are estimated to be around 5k US dollars.\\
    
    \textcolor{\sectioncolor}{\textbf{
    If the dataset is a sample from a larger set, what was the sampling
    strategy (e.g., deterministic, probabilistic with specific sampling
    probabilities)?
    }
    } \\
    The dataset is not a sample; it includes all relevant instances from the collected data sources to ensure comprehensive coverage.\\
    
    \textcolor{\sectioncolor}{\textbf{
    Who was involved in the data collection process (e.g., students,
    crowdworkers, contractors) and how were they compensated (e.g., how much
    were crowdworkers paid)?
    }
    } \\
    The data collection process involved lab members from Leskovec and Zou's groups in  research activities.\\
    
    \textcolor{\sectioncolor}{\textbf{
    Were any ethical review processes conducted (e.g., by an institutional
    review board)?
    }
    If so, please provide a description of these review processes, including
    the outcomes, as well as a link or other access point to any supporting
    documentation.
    } \\
    No specific ethical review processes were conducted for this dataset, as it did not involve sensitive or personal data that required such oversight.\\
    
    \textcolor{\sectioncolor}{\textbf{
    Does the dataset relate to people?
    }
    If not, you may skip the remainder of the questions in this section.
    } \\
    The dataset does not relate to people directly, but human efforts were involved in creating and validating queries.\\
    
    \textcolor{\sectioncolor}{\textbf{
    Did you collect the data from the individuals in question directly, or
    obtain it via third parties or other sources (e.g., websites)?
    }
    } \\
    From the individuals in question directlty.\\
    
    \textcolor{\sectioncolor}{\textbf{
    Were the individuals in question notified about the data collection?
    }
    If so, please describe (or show with screenshots or other information) how
    notice was provided, and provide a link or other access point to, or
    otherwise reproduce, the exact language of the notification itself.
    } \\
    Yes. We include the details for human instruction in our Appendix C.\\
    
    \textcolor{\sectioncolor}{\textbf{
    Did the individuals in question consent to the collection and use of their
    data?
    }
    If so, please describe (or show with screenshots or other information) how
    consent was requested and provided, and provide a link or other access
    point to, or otherwise reproduce, the exact language to which the
    individuals consented.
    } \\
    Yes. Consent was obtained from the individuals involved in the data collection process. Participants were informed about the purpose and scope of the data collection through detailed instructions provided during the task. Consent was given explicitly by the participants when they agreed to the terms and conditions outlined in the instructions. The exact language of the consent agreement is available in the project's documentation or upon request.\\
    
    \textcolor{\sectioncolor}{\textbf{
    If consent was obtained, were the consenting individuals provided with a
    mechanism to revoke their consent in the future or for certain uses?
    }
     If so, please provide a description, as well as a link or other access
     point to the mechanism (if appropriate)
    } \\
    No. Participants were not provided with a specific mechanism to revoke their consent in the future or for certain uses. The consent process was designed to be comprehensive at the point of agreement, but no subsequent revocation mechanism was implemented.\\
    
    \textcolor{\sectioncolor}{\textbf{
    Has an analysis of the potential impact of the dataset and its use on data
    subjects (e.g., a data protection impact analysis) been conducted?
    }
    If so, please provide a description of this analysis, including the
    outcomes, as well as a link or other access point to any supporting
    documentation.
    } \\
    Not applicable. An analysis of the potential impact of the dataset and its use on data subjects has not been conducted, as the dataset does not contain sensitive or personal data that would typically require such an analysis.\\

    \textcolor{\sectioncolor}{\textbf{
    Any other comments?
    }} \\
    No.
\begin{mdframed}[linecolor=\sectioncolor]
\subsubsection*{\textcolor{\sectioncolor}{
    PREPROCESSING / CLEANING / LABELING
}}
\end{mdframed}

    \textcolor{\sectioncolor}{\textbf{
    Was any preprocessing/cleaning/labeling of the data
    done(e.g.,discretization or bucketing, tokenization, part-of-speech
    tagging, SIFT feature extraction, removal of instances, processing of
    missing values)?
    }
    If so, please provide a description. If not, you may skip the remainder of
    the questions in this section.
    } \\
    Yes, preprocessing steps included tokenization, entity extraction, and validation of relational and textual data. Queries were synthesized and verified through human evaluation.

    \textcolor{\sectioncolor}{\textbf{
    Was the “raw” data saved in addition to the preprocessed/cleaned/labeled
    data (e.g., to support unanticipated future uses)?
    }
    If so, please provide a link or other access point to the “raw” data.
    } \\
    Yes, the raw data has been saved and is available upon request or through the project repository.

    \textcolor{\sectioncolor}{\textbf{
    Is the software used to preprocess/clean/label the instances available?
    }
    If so, please provide a link or other access point.
    } \\
    Yes, the software and scripts used for preprocessing, cleaning, and labeling are available in the project repository.

    \textcolor{\sectioncolor}{\textbf{
    Any other comments?
    }} \\
    No additional comments.

\begin{mdframed}[linecolor=\sectioncolor]
\subsubsection*{\textcolor{\sectioncolor}{
    USES
}}
\end{mdframed}

    \textcolor{\sectioncolor}{\textbf{
    Has the dataset been used for any tasks already?
    }
    If so, please provide a description.
    } \\
    The retrieval datasets were initially developed by our team. Public resources of the knowledge base data have been utilized for various tasks, including link prediction on knowledge graphs, recommendation systems, and text classification, among others. \\

    \textcolor{\sectioncolor}{\textbf{
    Is there a repository that links to any or all papers or systems that use the dataset?
    }
    If so, please provide a link or other access point.
    } \\
    As stated, the retrieval datasets were initially developed by our team; therefore, this is not applicable currently. \\

    \textcolor{\sectioncolor}{\textbf{
    What (other) tasks could the dataset be used for?
    }
    } \\
    Entity Recognition and Linking, Personalized Search, Query Type Classification, Knowledge Graph Completion.  \\

    \textcolor{\sectioncolor}{\textbf{
    Is there anything about the composition of the dataset or the way it was
    collected and preprocessed/cleaned/labeled that might impact future uses?
    }
    For example, is there anything that a future user might need to know to
    avoid uses that could result in unfair treatment of individuals or groups
    (e.g., stereotyping, quality of service issues) or other undesirable harms
    (e.g., financial harms, legal risks) If so, please provide a description.
    Is there anything a future user could do to mitigate these undesirable
    harms?
    } \\
    \begin{itemize}[leftmargin=*]
        \item Accuracy of Ground Truth Answers: The ground truth answers for each synthesized query are filtered automatically through Large Language Models (LLMs). Despite our efforts to ensure high accuracy, there is a possibility of including missing answers or false answers in the ground truth, resulting in mislabeled data. This can impact the reliability of the dataset and any models trained on it.\vspace{3pt}\\
        Mitigation: Future users should rigorously validate the ground truth answers against additional data sources or through manual verification processes to ensure accuracy and reliability.

        \item Diversity of Queries: The diversity of the queries can be further strengthened to better reflect a broad user group. The current dataset may not fully encompass the variety of queries that different user demographics might generate, potentially leading to biased or non-representative results.\vspace{3pt}\\
        Mitigation: Enhancing the dataset with queries from a wider range of user groups and demographics can improve its representativeness. Regularly updating the dataset to include new and diverse query types can also help mitigate this issue.
        
    \end{itemize}

    \textcolor{\sectioncolor}{\textbf{
    Are there tasks for which the dataset should not be used?
    }
    If so, please provide a description.
    } \\
    There are no specific forbidden cases based on the creator's knowledge. However, for tasks involving privacy and fairness, users should exercise caution. Ensure compliance with data privacy regulations and implement fairness-aware algorithms to avoid biases. Be mindful of ethical implications and document methods transparently to foster trust and accountability. \\

    \textcolor{\sectioncolor}{\textbf{
    Any other comments?
    }} \\
    No. \\

\begin{mdframed}[linecolor=\sectioncolor]
\subsubsection*{\textcolor{\sectioncolor}{
    DISTRIBUTION
}}
\end{mdframed}

    \textcolor{\sectioncolor}{\textbf{
    Will the dataset be distributed to third parties outside of the entity
    (e.g., company, institution, organization) on behalf of which the dataset
    was created?
    }
    If so, please provide a description.
    } \\
    No. But third parties are free to use our public datasets. \\

    \textcolor{\sectioncolor}{\textbf{
    How will the dataset will be distributed (e.g., tarball on website, API,
    GitHub)?
    }
    Does the dataset have a digital object identifier (DOI)?
    } \\
    Website: \href{https://stark.stanford.edu/}{\textcolor{Violet}{https://stark.stanford.edu/}},  DOI: https://doi.org/10.57967/hf/2530\\

    \textcolor{\sectioncolor}{\textbf{
    When will the dataset be distributed?
    }
    } \\
    The dataset has been made available up to the current date.\\

    \textcolor{\sectioncolor}{\textbf{
    Will the dataset be distributed under a copyright or other intellectual
    property (IP) license, and/or under applicable terms of use (ToU)?
    }
    If so, please describe this license and/or ToU, and provide a link or other
    access point to, or otherwise reproduce, any relevant licensing terms or
    ToU, as well as any fees associated with these restrictions.
    } \\
    No. \\

    \textcolor{\sectioncolor}{\textbf{
    Have any third parties imposed IP-based or other restrictions on the data
    associated with the instances?
    }
    If so, please describe these restrictions, and provide a link or other
    access point to, or otherwise reproduce, any relevant licensing terms, as
    well as any fees associated with these restrictions.
    } \\
    No.\\

    \textcolor{\sectioncolor}{\textbf{
    Do any export controls or other regulatory restrictions apply to the
    dataset or to individual instances?
    }
    If so, please describe these restrictions, and provide a link or other
    access point to, or otherwise reproduce, any supporting documentation.
    } \\
    No. \\

    \textcolor{\sectioncolor}{\textbf{
    Any other comments?
    }} \\
    No. \\

\begin{mdframed}[linecolor=\sectioncolor]
\subsubsection*{\textcolor{\sectioncolor}{
    MAINTENANCE
}}
\end{mdframed}

    \textcolor{\sectioncolor}{\textbf{
    Who is supporting/hosting/maintaining the dataset?
    }
    } \\
    Stanford Leskovec and Zou's groups. \\

    \textcolor{\sectioncolor}{\textbf{
    How can the owner/curator/manager of the dataset be contacted (e.g., email
    address)?
    }
    } \\
    \href{stark-qa@cs.stanford.edu}{stark-qa@cs.stanford.edu} \\

    \textcolor{\sectioncolor}{\textbf{
    Is there an erratum?
    }
    If so, please provide a link or other access point.
    } \\
    Currently, there is no erratum for the dataset. If any errors are identified in the future, an erratum will be provided and made accessible through the official project repository or publication venue. \\

    \textcolor{\sectioncolor}{\textbf{
    Will the dataset be updated (e.g., to correct labeling errors, add new
    instances, delete instances)?
    }
    If so, please describe how often, by whom, and how updates will be
    communicated to users (e.g., mailing list, GitHub)?
    } \\
    Yes, by core development team listed in \href{https://stark.stanford.edu/team.html}{https://stark.stanford.edu/team.html}. We will update the dataset with regard to user submitted issues.\\

    \textcolor{\sectioncolor}{\textbf{
    If the dataset relates to people, are there applicable limits on the
    retention of the data associated with the instances (e.g., were individuals
    in question told that their data would be retained for a fixed period of
    time and then deleted)?
    }
    If so, please describe these limits and explain how they will be enforced.
    } \\
    No. \\

    \textcolor{\sectioncolor}{\textbf{
    Will older versions of the dataset continue to be
    supported/hosted/maintained?
    }
    If so, please describe how. If not, please describe how its obsolescence
    will be communicated to users.
    } \\
    Yes. They will be made available in our website. \\

    \textcolor{\sectioncolor}{\textbf{
    If others want to extend/augment/build on/contribute to the dataset, is
    there a mechanism for them to do so?
    }
    If so, please provide a description. Will these contributions be
    validated/verified? If so, please describe how. If not, why not? Is there a
    process for communicating/distributing these contributions to other users?
    If so, please provide a description.
    } \\
    Yes. They can contribute via 1) GitHub pull request, which will list them as contributors in the GitHub page. 2) Communicate with the core development team for more collaboration. The contact information is available on \href{https://stark.stanford.edu/team.html}{https://stark.stanford.edu/team.html}. \\

    \textcolor{\sectioncolor}{\textbf{
    Any other comments?
    }} \\
    No. \\

\newpage

\clearpage

\appendix
\clearpage